\documentclass[aps,prd,twocolumn,groupedaddress]{revtex4-1}
\usepackage{amsmath}
\usepackage{amsfonts}
\usepackage{amssymb}
\usepackage{graphicx}
\usepackage{subcaption}
\usepackage{color}
\usepackage[dvipsnames]{xcolor}
\usepackage[bookmarks=true,colorlinks=true]{hyperref}
\hypersetup{linkcolor=blue}
\hypersetup{citecolor=blue}
\hypersetup{urlcolor=blue}

\begin{document}


\title{ \bf Non-critical anisotropic Bianchi type $I$ string cosmology with $\alpha'$-corrections}


\author{ F. Naderi}
\email[]{f.naderi@azaruniv.ac.ir}
\affiliation{Young Researchers and Elite Club, Marand Branch, Islamic Azad university, Marand, Iran}
\author{A. Rezaei-Aghdam}
\email[]{rezaei-a@azaruniv.ac.ir}
\affiliation{Department of Physics, Faculty of Basic Sciences, Azarbaijan Shahid Madani University, 53714-161, Tabriz, Iran}
\author{ F. Darabi}
\email[]{f.darabi@azaruniv.ac.ir}
\affiliation{Department of Physics, Faculty of Basic Sciences, Azarbaijan Shahid Madani University, 53714-161, Tabriz, Iran}

\date{\today}

\begin{abstract}
We present non-critical  Bianchi type $I$  string cosmology solutions in the presence of central charge deficit term $\Lambda$. The leading order string frame curvature appears to be in the high curvature limit $R\alpha'\gtrsim1$,  which underlines the necessity of including higher order $\alpha'$-corrections.    
We give new solutions of two-loop (first order $\alpha'$) $\beta$-function equations of  $\sigma$-model with non-zero $\Lambda$ and dilaton field in both  cases of absence and presence of spatially homogeneous $H$-field ($H=dB$). Also, the evolution of solutions is studied in the Einstein frame, where the string effective action can transform to  Gauss-Bonnet gravity model coupled to  dilaton field with potential. We study  explicit examples in first order $\alpha'$ with
chosen values of appeared constants in the solutions and discuss the cosmological implications.
\end{abstract}

\pacs{}

\maketitle

\section{Introduction}
In  $\sigma$-model context, the conformal invariance   is provided   by  
vanishing  $\beta$-functions \cite{TSEYTLIN198634}, which are equivalent to   the equations of motion
of  effective action in  string frame \cite{metsaev1987order}.
The  low energy   string effective action, being compatible with the conformal invariance in one-loop order,  has  wide cosmological implication for describing the evolution of  early  universe 
with a very low curvature and string coupling, $g= e^{-\phi}$ 
\cite{CALLAN1985593,fradkin1985quantum,gasperini1997towards}.
In  two-loop  order of $\beta$-functions,   the string  effective action is modified by including  the $\alpha'$-corrections of   quadratic
curvature type $\alpha'R^2$, where the $\alpha'$ is  square of  string-length,  $\alpha'=\lambda_s^2/2\pi$ \cite{metsaev1987order,HULL1988197}. The expanded leading order effective action is widely believed to regularize the curvature singularity  \cite{brustein1994graceful}.  The  two-loop $\beta$-functions,  possible  $\alpha'$-corrected  string effective actions and 
on-shell compatibility of  the $\alpha'$-corrected  effective action equations of motion  with the  two-loop conformal invariance condition have been investigated in \cite{metsaev1987order}.
A renormalization scheme (RS) dependence appears in the $B$-field dependent terms of two-loop $\beta$-functions and consequently in the $\alpha'$-corrected effective actions, where the two schemes of Gauss-Bonnet and $R^2$ have been considered distinctly in  \cite{metsaev1987order}. 
Furthermore,  the $S$-matrix is invariant under a set of field redefinitions \cite{Metsaev1987}, which allows to transform between RS and leads to a physically equivalent class of  effective actions    \cite{TSEYTLIN198692}. 

Generally,   two kinds of corrections can be included in string effective action,  the
stingy type 
$\alpha'$-expansion and the
quantum nature loop expansion in string coupling \cite{gasperini1997towards}.
The $\alpha'$-corrections are significant when the curvature is in the high limit $R\alpha'\gtrsim 1$, while the loop corrections become important in the case of strong string coupling $g_s>1$. As long as the coupling is sufficiently  weak in the high curvature regime, the $\alpha'$-corrections are enough to be taken into account and the loop corrections can be neglected  \cite{gasperini1997towards}.

Solutions of one-loop $\beta$-function equations  with contribution of dilaton
field and antisymmetric $B$-field have been presented for   several cosmological backgrounds such as homogeneous anisotropic space-times  \cite{Batakis1,batakis2,batakis3,barrow1997spatially,barrow1997kantowski} and   inhomogeneous models  \cite{barrow1997inhomogeneous}. 
According to \cite{batakis3},  the contribution of field strength tensor $H=dB$ in all Bianchi type  models with diagonal  metrics can be
classified into three classes of $\chi(\rightarrow)$,  
$\chi(\uparrow)$ and $\chi(\nearrow)$, where the $\chi$ is all possible Bianchi types and the arrows indicates the
orientation of     $H^*$, the dual of $H$ with respect to the 3D hypersurface of homogeneity  $\Sigma^3$ sections. 
In aforementioned works, the central charge
deficit term $\Lambda$ has been considered to be zero. In $D$-dimensions, 
$\Lambda$  is proportional to $D-26$ in bosonic string theory and $D-10$ in superstring theory and provides a term in the effective action analogous to the non-vanishing cosmological constant term \cite{tseytlin1992elements}. 
Solutions  with non-zero $\Lambda$, called non-critical string cosmology, have been obtained in  the lowest order $\beta$-function equations  \cite{tseytlin1992elements,0264-9381-9-4-013,PhysRevD.49.5019}.

Moreover, neglecting the $\Lambda$ term, 
the solutions of two-loop   $\beta$-function equations  have been presented  
in some works,   such as on the Kasner and Schwarzschild background with setting $H$-field to zero \cite{Exirifard}, and on anisotropic homogeneous backgrounds with the contribution of $H$-field  \cite{NADERI2017}.
Alternatively,
the   $\alpha'$-corrected field equations of effective action in Einstein frame have been solved in 
several  classes of backgrounds such as  M-theory, black hole, and cosmology   with no contribution of  $H$-field 
\cite{niz2007stringy,KawaiI,KawaiIx,tsujikawa2006cosmologies,cartier2001gravitational,kanti1999singularity,BF2,Bf,barrow1998godel}. Also, the solutions in the presence of $H$-field have been investigated for instance on FRW and GHS black hole backgrounds with zero  $\Lambda$  in \cite{BF2,Bf}, and  on Godel space-time with contribution of  $\Lambda$ in \cite{barrow1998godel}.

{  
	Especially,  attempts to find      accelerated expanding universes in the context of higher-dimensional superstring and $M$-theory led the people to consider the extended gravitational actions, since, in the low energy limit of their effective field theory, where the gravitational action is given only by Einstein-Hilbert action, the accelerated expanding solutions are not allowed with a time-independent    internal space   \cite{maldacena2001supergravity}.
	In these theories,  inflation is expected to occur at  Planck scale of ten or eleven dimensions, and in such a high energy scale the higher order corrections are required to be taken into account,  at least in the early times. In this sense, accelerated solutions    have been found in higher order corrected high dimensional string and $M$-theory  in the absence of $H$-field,   for example, in        \cite{maeda1988inflation,maeda2004inflation,bamba2007accelerating,akune2006inflation,maeda2012accelerating},
	with an especial attention to the de-sitter like and power-law expanding solutions.}

In this work, aimed at presenting a non-critical $4$-dimensional  two-loop string cosmology,  we study the solutions of two-loop $\beta$-function equations  on anisotropic Bianchi type $I$ space-time  with non-zero $\Lambda$ and dilaton field in two cases of the presence and absence of $H$-field. {As we will show, the leading order solutions have string frame curvature in the high limit $R\alpha'\gtrsim 1$, where the higher order $\alpha'$-corrections become significant. We will limit our calculations to the first order in $\alpha'$, where the corrections of quadratic curvature types are included in the effective action. 
	Concerning only this order of corrections, 
	the regularizing effects of $\alpha'$-corrections are already known \cite{gasperini1997towards,cartier2001gravitational,niz2007stringy,BF2}, where the higher order corrections usually act  to correct the lower order solutions order by order. We can, therefore, hope to provide a glimpse of the feature that could be obtained considering all orders in $\alpha'$.} Similar to what we have done in \cite{NADERI2017} for classifying and solving the two-loop $\beta$-function     on all Bianchi type models with $\Lambda=0$ and $H$-field in $\chi(\uparrow)$ class, a perturbative series expansion in $\alpha'$ is implemented on background field and the general forms of equations and solutions are presented.  Also,  we consider the field equations in  Einstein frame by obtaining the contribution of $H$-field in $\alpha'$ order of energy-momentum tensor  to investigate the cosmological implications of the $\alpha'$-corrected  solutions.

The paper is organized as follows. In section \ref{sec2}, the general forms of two-loop $\beta$-functions considering the two RS of Gauss-Bonnet and $R^2$ are recalled. Also, the field equations  of the Gauss-Bonnet scheme in the Einstein frame is presented. In section \ref{sec3}, the   two-loop $\beta$-functions with non-zero $\Lambda$ are solved on Bianchi type $I$ background in two cases of vanishing and non-vanishing $H$-field. Then, the behavior of solutions is investigated in the Einstein frame in section \ref{sec4}. Finally,    the main results are summarized in section \ref{Conclusion}.

\section{Two-loop (order $\alpha'$) $\beta$-functions and $\alpha'$-corrected string effective action}\label{sec2}

In a $\sigma$-model  with  background fields of metric $g$, dilaton field
$\phi$, and antisymmetric $B$-field,  the two-loop $\beta$-function of metric   
is given by \cite{metsaev1987order}
\begin{equation}\label{betaGR}
\begin{aligned}
\frac{1}{\alpha'}{\beta}^{g}_{\mu\nu}=&{R}_{{\mu}{\nu}}-\frac{1}{4}{H}^{2}_{{\mu}{\nu}}-\nabla_{{\mu}}\nabla_{{\nu}}{\phi}+\frac{\alpha'}{2}\big[
R_{\mu\alpha\beta\gamma}R_{\nu}^{~\alpha\beta\gamma}
\\
&-\frac{3}{2}R_{(\mu}^{~~~\alpha\beta\gamma}H_{\nu)\alpha\lambda}H_{\beta\gamma}^{~~\lambda}
-\frac{1}{2}R^{\alpha\beta\rho\sigma}H_{\mu\alpha\beta}H_{\nu\rho\sigma}         \\
&+\frac{1}{8}(H^4)_{\mu\nu}
-\frac{f}{2}\big(R_{\mu\alpha\beta\nu}(H^2)^{\alpha\beta}
 \\
&+2\,R_{(\mu}^{~~\alpha\beta\gamma}H_{\nu)\alpha\lambda}H_{\beta\gamma}^{~~\lambda}
+R^{\alpha\beta\rho\sigma}H_{\mu\alpha\beta}H^{\nu\rho\sigma}
 \\
&-\nabla_{\lambda}H_{\mu\alpha\beta}\nabla^{\lambda}H_{\nu}^{~\alpha\beta}
\big)   -\frac{1}{12}\nabla_{\mu}\nabla_{\nu}H^2\\
&
+\frac{1}{4}\nabla_{\lambda}H_{\mu\alpha\beta}\nabla^{\lambda}H_{\nu}^{~\alpha\beta}
+\frac{1}{12}\nabla_{\mu}H_{\alpha\beta\gamma}\nabla_{\nu}H^{\alpha\beta\gamma}
 \\
&+\frac{1}{8}H_{\mu\alpha\lambda}H_{\nu\beta}^{~~\lambda}(H^2)^{\alpha\beta}\big],
\end{aligned}
\end{equation}
where  $H^4=H_{\mu\nu\lambda}H^{\nu\rho\kappa}H_{\rho\sigma}^{~~\lambda}H^{\sigma\mu}_{~~\kappa}$,  $H_{\mu\nu}^2=H_{\mu\rho\sigma}H_{\nu}^{\rho\sigma}$,   and $H$ is  the field strength of $B$-field defined by $H_{\mu\nu\rho}=3\partial_{[\mu}B_{\nu\rho]}$. The $f$ parameter indicates the RS dependence in $\beta$-functions. 
Especially, the corresponding schemes to  $f=1$ and $f=-1$, called $R^2$ and {Gauss-Bonnet} schemes,    have been pointed in \cite{metsaev1987order}. Solutions of various RS $\beta$-function equations are different, but still equivalent because of their belonging to various definitions of physical metric, dilaton field, and $B$-field.
The $\beta$-functions of $B$-field in mentioned RS are given by \cite{metsaev1987order}
\begin{eqnarray}\label{betaBR}
\begin{split}
\frac{1}{\alpha'}{{\beta}}_{{\mu}{\nu}}^{{B}}(f&=1)=-\frac{1}{2}\nabla^{{\mu}}{H}_{{\mu}{\nu}{\rho}}+\frac{\alpha'}{4}(2R_{[\mu\gamma\alpha\beta}\nabla^{\gamma}H^{\alpha\beta}_{~~~\nu]} \\
&+\nabla_{\gamma}H_{\alpha\beta[\mu}H_{\nu]\rho}^{~~~\alpha}H^{\beta\gamma\rho}+2\nabla_{\beta}H^2_{\alpha[\nu}H_{\mu]}^{~~\alpha\beta} \\
&+\frac{1}{2}H^2_{\alpha\beta}\nabla^{\alpha}H^{\beta}_{~\mu\nu}-\frac{1}{12}H_{\mu\nu}^{~~\rho}\nabla_{\rho}H^2),
\end{split}
\end{eqnarray}
\begin{equation}\label{betaBG}
\begin{aligned}
\frac{1}{\alpha'}{\beta}^{B}_{\mu\nu}(f&=-1)=\hat{R}_{[{\mu}{\nu}]}+\frac{\alpha'}{4}\big(2\hat{R}^{\alpha\beta\gamma}_{~~~~[\nu}\hat{R}_{\mu]\alpha\beta\gamma}\\
&-\hat{R}^{\beta\gamma\alpha}_{~~~~[\nu}\hat{R}_{\mu]\alpha\beta\gamma}+\hat{R}_{\alpha[\mu\nu]\beta}H^{\alpha\beta}-\frac{1}{12}H_{\mu\nu}^{~~\rho}\nabla_{\rho}H^2\big),
\end{aligned}
\end{equation}
where   the $\hat{R}^{\rho}_{\mu\nu\sigma}$ is the Riemann tensor of the generalized connection with torsion ${\hat{\Gamma}}^{\rho}_{\mu\nu}={{\Gamma}}^{\rho}_{\mu\nu}-\frac{1}{2}H^{\rho}_{\mu\nu}$ \footnote{
	${\hat{R}}^{\kappa}_{~\lambda\mu\nu}= R^{\kappa}_{~\lambda\mu\nu}-\frac{1}{2}\nabla_{\mu}H^{\kappa}_{\nu\lambda}-\frac{1}{2}\nabla_{\nu}H^{\kappa}_{\mu\lambda}+\frac{1}{4}H_{~\nu\lambda}^{\gamma}H_{~\mu\gamma}^{\kappa}-\frac{1}{4}H_{~\mu\lambda}^{\gamma}H^{\kappa}_{~\nu\gamma}
	$}.
The averaged    $\beta$-function of dilaton, which can be written in terms of $\beta$-functions of metric and dilaton field  as $   \tilde{\beta}^{{\phi}}={\beta}^{{\phi}}-\frac{1}{4}{\beta}^{g}_{\mu\nu}g^{\mu\nu}$, is given  by
\begin{eqnarray}\label{betafR}
\begin{split}
\frac{1}{\alpha'}\tilde{\beta}^{{\phi}}&=-{R}+\frac{1}{12}{H}^{2}+2\nabla_{{\mu}}\nabla^{{\mu}}{\phi}+(\partial_{{\mu}}{\phi})^{2}-\Lambda-\frac{\alpha'}{4}(R^2_{\mu\nu\rho\lambda}\\
&-\frac{1}{2}R^{\alpha\beta\rho\sigma}H_{\alpha\beta\lambda}H_{\rho\sigma}^{~~\lambda}
+\frac{1}{24}H^4-\frac{1}{8}(H_{\mu\nu}^{~~2})^2),
\end{split}
\end{eqnarray}
which can be obtained   
by variation of the following string effective action with respect to the dilaton field
\begin{eqnarray}\label{action}
\begin{split}
S=\int &d^4 x\sqrt{G}e^{\phi}(R-\frac{1}{12}H^2+(\nabla\phi)^2+\Lambda+\frac{\alpha'}{4}(R^2_{\mu\nu\rho\lambda}\\
&-\frac{1}{2}R^{\alpha\beta\rho\sigma}H_{\alpha\beta\lambda}H_{\rho\sigma}^{~~\lambda}
+\frac{1}{24}H^4-\frac{1}{8}(H_{\mu\nu}^{~~2})^2)).
\end{split}
\end{eqnarray}  
The $\Lambda$ term is related to the central charge deficit of theory and in non-critical $D$-dimensional bosonic theory is  given  by \cite{0264-9381-9-4-013,tseytlin1992elements}
\begin{equation}        
\Lambda=\frac{2\,(26-D)}{3\alpha'}.
\end{equation} 
The effective action \eqref{action} has been written in string frame and its variations with respect to the background fields give the $\beta$-functions. Also, there is another frame,   namely  the   Einstein frame which its metric, $\tilde{g}_{\mu\nu}$,  is related to the string frame metric, ${g}_{\mu\nu}$, in $4$-dimensional space-time by
\begin{equation}\label{3}
\tilde{ g}_{\mu\nu}=e^{\phi}g_{\mu\nu}.
\end{equation}   
Actually, the $g_{\mu\nu}$ is the metric seen by the string and describes physics from the string viewpoint. However,  it is not convenient to understand the gravitational phenomena due to the dilaton field dependent coefficient of Ricci scalar in \eqref{action}. Transforming to Einstein frame by performing the conformal transformation \eqref{3} eliminates the dilaton field dependent factor.
{This frame is appropriate for comparison with the string $S$-matrix. Actually, computing the $\alpha'$-corrected string effective action can be studied either in the $\sigma$-model and its $\beta$-functions approach or from the tree-level $S$-matrix. However, it is worth noting that to a given order $\alpha'$ an intrinsic ambiguity remains in the string effective actions. 
	Since the  $S$-matrix  is invariant under  a set of field redefinitions of type \cite{Metsaev1987}
   \begin{equation*}\label{mm}
\begin{split}
\delta g_{\mu\nu}=&\alpha'(b_1R_{\mu\nu}+b_2\partial_{\mu}\phi\partial_{\nu}\phi+b_3H^2_{\mu\nu}+g_{\mu\nu}(b_4R\\
&~~~~+b_5(\partial \phi)^2+b_6\nabla^2\phi+b_7H^2)),\\
\delta B_{\mu\nu}=&\alpha'(b_8\nabla^{\lambda}H_{\lambda\mu\nu}+b_9H_{\mu\nu}^{~~\lambda}\partial_{\lambda}\phi),\\
\delta \phi=&\alpha'(b_{10}R+b_{11}(\partial\phi)^2+b_{12}\nabla^2\phi+b_{13}H^2),
\end{split}
\end{equation*}
 there is a field redefinition ambiguity and a class of physically equivalent effective actions parametrized by $8$   essential coefficients  \cite{Tseytlin2}. Choosing a particular set of field variables corresponds to a particular  RS choice.  Aimed at calculating a set of these coefficients,  the Gauss-Bonnet scheme has been used  which gives the following effective action for the bosonic string in $4$ dimensions \cite{metsaev1987order} }

\begin{eqnarray}\label{GBaction}
\begin{split}
S=&\int d^4 x\sqrt{G}\big(\tilde{R}-\frac{1}{12}\,e^{2\phi}H^2-\frac{1}{2}(\tilde{\nabla}\phi)^2+\Lambda e^{-\phi}\\
&+\frac{\alpha'e^{\phi}}{4}\big[\tilde{R}^2_{\mu\nu\rho\lambda}-4\,\tilde{R}_{\mu\nu}^2+\tilde{R}^2+e^{\phi}(\frac{1}{2}H^2_{\mu\nu}\tilde{\nabla}^{\mu}\phi\tilde{\nabla}^{\nu}\phi\\
&-\frac{1}{2}\tilde{R}^{\alpha\beta\rho\sigma}H_{\alpha\beta\lambda}H_{\rho\sigma}^{~~\lambda}
+\frac{1}{2}H^2_{\mu\nu}\tilde{\nabla}^{\mu}\phi\tilde{\nabla}^{\nu}\phi-\frac{1}{12}H^2(\tilde{\nabla}\phi)^2)\\
&+e^{2\phi}(\frac{1}{24}H^4+\frac{1}{8}(H_{\mu\nu}^{~~2})^2-\frac{5}{144}(H^2)^2\big)]\big),
\end{split}
\end{eqnarray}
in which  $\tilde{\nabla}$ indicates the covariant derivative with respect to  $\tilde{g}$.  The $\Lambda$, which is positive in $D=4$,  appeared in a way reminding a negative  cosmological constant in  standard
theory of gravity  but along with a weight factor $e^{-\phi}$.  
{Using the field redefinitions, the Gauss-Bonnet combination  $\tilde{R}^2_{\mu\nu\rho\lambda}-4\,\tilde{R}_{\mu\nu}^2+\tilde{R}^2$ in the effective action can be replaced by the square of the Riemann tensor at the price of appearing a dilaton-dependent $\alpha'$-correction \cite{metsaev1987order,HULL1988197}.}

The equivalence of two-loop $\beta$-functions and equations of motion of $\alpha'$-corrected effective action can be established by using the field redefinitions  {and the lowest order equations of motion} \cite{metsaev1987order}. Physical quantities are not affected by the field redefinitions \cite{Bf}, and appropriate using of them and the leading order equations of motions  allows to transform between the RSs 
\cite{metsaev1987order,Metsaev1987}.
Hence,  where the higher order corrected field equations of effective actions are considered in the string theory context, the field redefinitions can be applied conveniently to reach the simplest effective action. 
In this sense,
the Gauss-Bonnet effective action usually holds attention which  is free of ghost and terms with higher than the second derivative in its field equations. Considering the effective action \eqref{GBaction}, the variation over the Einstein frame metric metric, $\tilde{    g}_{\mu\nu}$,  gives
\begin{eqnarray}\label{22}
\begin{split}
\tilde{R}_{\mu\nu}-\frac{1}{2}       \tilde{R}       \tilde{g}_{\mu\nu}=T_{\mu\nu}^{\mathrm{(eff)}},
\end{split}
\end{eqnarray}
where the effective  energy-momentum tensor is defined as follows 
\begin{eqnarray}\label{Teff}
\begin{split}
T_{\mu\nu}^{\mathrm{(eff)}}   =T_{\mu\nu}^{(\phi)}+T_{\mu\nu}^{(B_1)}+T_{\mu\nu}^{(\mathrm{GB})}+T_{\mu\nu}^{{(B_2)}},
\end{split}
\end{eqnarray}
in which the energy-momentum tensors of dilaton field and $B$-field  in the leading order are given by
\begin{eqnarray}
T_{\mu\nu}^{(\phi)}=\frac{1}{2}(\tilde{\nabla}_{\mu}\phi\tilde{\nabla}_{\nu}\phi-\frac{1}{2}\,\tilde{g}_{\mu\nu}(\tilde{\nabla}\phi)^2+\Lambda\,e^{\phi}\tilde{g}_{\mu\nu}),\label{Tfi}\\
T^{(B_1)}_{\mu\nu}=\frac{e^{2\,\phi}}{4}(H_{\mu\kappa\lambda}H_{\nu}^{\kappa\lambda}-\frac{1}{6}H^2\tilde{g}_{\mu\nu})\label{TB},
\end{eqnarray}
	and in the $\alpha'$ order,  the Gauss-Bonnet term gives \cite{nojiri2005gauss}
\begin{widetext}
	\begin{eqnarray}\label{TGB}
	\begin{split}
	T_{\mu\nu}^{(\mathrm{GB})}&=\alpha'e^{\phi}\big[-\frac{1}{2} \tilde{R}_{\mu\alpha\beta\gamma}        \tilde{R}_{\nu}^{~\alpha\beta\gamma}+\tilde{R}_{\mu\alpha\beta\nu}\tilde{R}^{\alpha\beta}+\tilde{R}_{\mu\alpha}\tilde{R}^{\nu\alpha}-\frac{1}{2}\tilde{R}\tilde{R}_{\mu\nu}+\frac{1}{8}\tilde{g}_{\mu\nu}(\tilde{R}^2_{\alpha\beta\rho\lambda}-4\,\tilde{R}_{\alpha\beta}^2+\tilde{R}^2)\\\
	&+\tilde{R}_{\mu\alpha\beta\nu}(\tilde{\nabla}^{\alpha}\tilde{\nabla}^{\beta}\phi+\tilde{\nabla}^{\alpha}\phi\tilde{\nabla}^{\beta}\phi)-2\tilde{R}_{~(\mu}^{\alpha}(\tilde{\nabla}_{\nu)}\tilde{\nabla}_{\alpha}\phi+\tilde{\nabla}_{\nu)}\phi\tilde{\nabla}_{\alpha}\phi)
	-\tilde{R}_{\mu\nu}((\tilde{\nabla}\phi)^2+\tilde{\nabla}^2\phi)\\
	&+\frac{1}{2}\tilde{R}(\tilde{\nabla}_{\mu}\phi\tilde{\nabla}_{\nu}\phi+\tilde{\nabla}_{\mu}\tilde{\nabla}_{\nu}\phi)+\big((\tilde{\nabla}_{\rho}\tilde{\nabla}_{\sigma}\phi+\tilde{\nabla}_{\rho}\phi\tilde{\nabla}_{\sigma}\phi)\,\tilde{R}^{\rho\sigma}-\frac{1}{2}((\tilde{\nabla}\phi)^2+\tilde{\nabla}^2\phi)\,\tilde{R}\big)\tilde{g}_{\mu\nu}\big].
	\end{split}
	\end{eqnarray}
	Also,  we obtain  the following energy-momentum tensor for  the $B$-field dependent $\alpha'$-correction terms
\begin{eqnarray}\label{TB2}
\begin{split}
T_{\mu\nu}^{(B_2)}&= \frac{\alpha'e^{3\,\phi}}{8}\big[\big(
\tilde{R}^{\alpha\beta\rho\sigma}H_{\mu\alpha\beta}H_{\nu\rho\sigma}
+3\tilde{R}^{\alpha\beta\sigma}_{~~~~(\mu}H_{\nu)\lambda\sigma}H_{~\alpha\beta}^{\lambda}
+e^{-3\,\phi}\tilde{\nabla}^{\rho}\tilde{\nabla}^{\sigma}(e^{3\,\phi} H_{\mu\rho\lambda}H_{\nu\sigma}^{~~\lambda})
-\frac{1}{2}(H_{\alpha\mu\sigma}H_{\beta\nu}^{~~\sigma}\tilde{\nabla}^{\alpha}\phi\tilde{\nabla}^{\beta}\phi\\
&+H^2_{\beta(\mu}\tilde{\nabla}_{\nu)}\phi\tilde{\nabla}^{\beta}\phi)+\frac{1}{2}(\tilde{\nabla}\phi)^2H^2_{\mu\nu}+\frac{1}{12}H^2\tilde{\nabla}_{\mu}\phi\tilde{\nabla}_{\nu}\phi
\big)-\frac{e^{2\,\phi}}{2}\big(H^4_{\mu\nu}+H^2_{\mu\rho}H_{\nu}^{2\,\rho}+2H^{2}_{\alpha\beta}H_{\mu}^{~\alpha\lambda}H_{\nu\lambda}^{~~\beta}-\frac{5}{12}H^2H^2_{\mu\nu}\big)\\
&+\frac{1}{2}\tilde{g}_{\mu\nu}\big(\big(-\tilde{R}^{\alpha\beta\rho\sigma}H_{\alpha\beta\gamma}H_{\rho\sigma}^{~~\gamma}-\tilde{\nabla}^{\rho}\phi\tilde{\nabla}^{\beta}\phi H^2_{\rho\beta}-\frac{1}{6}H^2(\tilde{\nabla}\phi)^2\big)
+{e^{2\,\phi}}\big(\frac{1}{12}H^4+\frac{1}{4}(H_{\mu\nu})^2-\frac{5}{72}(H^2)^2\big)
\big)
\big].
\end{split}
\end{eqnarray}
\end{widetext}
The $T_{\mu\nu}^{\mathrm{GB}}$ in \eqref{TGB} has been written in its most general case.  Indeed,   the Gauss-Bonnet term is a total derivative in $4$-dimensions and   the terms  without derivatives of dilaton in $T_{\mu\nu}^{\mathrm{GB}}$  cancel each other and vanish automatically \cite{nojiri2005gauss}.

\section{Non-critical ($\Lambda\neq 0$) anisotropic Bianchi type $I$ two-loop string cosmology solutions } \label{sec3}
In this section, we are going to solve the two-loop  $\beta$-functions in the presence of central charge deficit $\Lambda$ on Bianchi-type $I$ space-time with a similar method by which  we have calculated  the solutions of two-loop $\beta$-functions on homogeneous space-times in \cite{NADERI2017}.  The solutions give the string frame metric $g_{\mu\nu}$, dilaton field, and $B$-field, where the corresponding  Einstein frame solutions can be obtained using the conformal transformation \eqref{3}.
Maintaining   the provided convention by $\beta$-function solutions, 
the field redefinitions of \cite{Metsaev1987} will not be applied.    

Considering the  
anisotropic Bianchi-type $I$ metric as string frame  metric
\begin{eqnarray}\label{metric}
ds^2=g_{\mu\nu}\,dx^{\mu}dx^{\nu}=-g_{00}(t)dt^{2}+\sum_{i=1}^{3} a_i^2(t)(dx^{i})^2
\end{eqnarray}
where $a_i$ are the string frame scale factors,  we have
\begin{eqnarray}
\begin{split}
&\Gamma^{t}_{ij}=g^{00} a_i(t)^2H_i \, \delta_{j}^{i} ,\quad  \Gamma^{i}_{jt}=H_i\delta_{j}^{i},\quad \Gamma^{t}_{tt}=H_{0},\label{k1}\\
&R_{ijij}=g^{00} (a_i\,a_j)^2H_i\,H_j, \quad R_{itit}=-a_i^2(\dot{H_i}-H_iH_0+H_i^2),\\
&R_{tt}=-\sum (\dot{H_i}+H_i^2-H_0\,H_i),\\
& R_{ij}=\dot{H_i}+H_i\sum H_k-H_0\,H_i,\\
&R=g^{00}(\sum (2\,\dot{H_i}+H_i^2-2\,H_0\,H_i)+(\sum H_i)^2). \label{k2}
\end{split}
\end{eqnarray}
The  dot symbol stands for derivation with respect to $t$ and the  $H_i$ are  the Hubble coefficients of string frame  defined by 
$H_{i}={\frac{d}{d\,t}(\ln a_{i})}$ and $H_{0}=\frac{1}{2}{\frac{d}{d\,t}(\ln g_{00})}$. 

The solutions of $\beta$-function equations will be investigated in  two cases of absence and presence of $H$-field. Since the considered metric is spatially homogeneous, the dilaton field
can be only a function of time.
\subsection{Solutions with vanishing  $H$-field }\label{H=0}
Without contribution of  $H$-field,  the $(i,i)$ and time-time components of metric $\beta$-function \eqref{betaGR} and the $\beta$-function of dilaton \eqref{betafR}  
with using  relations \eqref{k1} reduce the following coupled differential equations  
\begin{eqnarray}\label{ii}
\dot{H_{i}}+ H_i{\sum}H_{k} + H_{i} \dot{\phi}-H_{i}\,H_0+\alpha'K_{i}g^{00}=0,
\end{eqnarray}
\begin{eqnarray}\label{00}
\ddot{\phi}+{\sum}(\dot{H_i}+ H_i^2-H_0(\dot{\phi}+\dot{H_i}))+\alpha'K_{0}g^{00}=0,
\end{eqnarray}
\begin{eqnarray}\label{fi}
\begin{aligned}
-2\ddot{\phi}&-\dot{\phi}^2-{\sum}(2\,\dot{\phi} H_i+{H_i}^2+2\,\dot{H_i})
- ({\sum} H_i)^2-\Lambda g_{00}\\
&-2(\phi'+\sum H_i)H_0
-\alpha'K_{\phi}g^{00}=0,
\end{aligned}
\end{eqnarray}
where the auxiliary $K$ functions have been introduced for  the shorthanded writing of equations and are given as follows
\begin{eqnarray}\label{kii}
\begin{aligned}
&K_{i}=\dot{H^2_{i}}+2(H_{i}-H_0)H_i\dot{H_{i}}+H^2_{i}(\sum H^2_{k}-H_0^2-H_0H_i),\\
&K_{0}=\sum(\dot{H^2_{i}}+2H_i(H_i-H_0)\dot{H_{i}}+H^4_{i}+H_0^2 H_i^2-H_0 H_i^3),\\
&K_{\phi}=
\sum(\dot{H^2_{i}}+2H_i(H_i-H_0)\dot{H_{i}}+{H^4_{i}}\\
&~~~~~~~~~~~~~~~~~~+H_0^2 H_i^2-H_0 H_i^3)+\sum_{i<j}H_i^2H_j^2.
\end{aligned}
\end{eqnarray}
Now, adding the  summed over $i$ of \eqref{ii} and  $\eqref{00}$ to  \eqref{fi} leads to the following equation
\begin{eqnarray}\label{33}
\begin{aligned}
-\ddot{\phi}-\dot{\phi}(\dot{\phi}&+\sum H_i)+\dot{\phi}H_0-\Lambda\,g_{00} \\
&+\alpha'(K_{\phi}+K_{0}+\sum K_i)g^{00}=0.
\end{aligned}
\end{eqnarray}
Also, subtracting the summed over $i$ of
\eqref{ii} from the sum of equations \eqref{00} and \eqref{33}   gives the initial value equation as follows
\begin{eqnarray}\label{ini}
\begin{aligned}
\dot{\phi}(\dot{\phi}+2\sum H_i)+(\sum H_i)^2&-\sum H_i^2+\Lambda\,g_{00} \\
&-\alpha'(K_{\phi}+2K_0)g^{00}=0.
\end{aligned}
\end{eqnarray}
We are going to solve the set of equations of \eqref{ii} and \eqref{33} subject to the  initial value equation \eqref{ini} along with implementing the following perturbative series expansion on the background fields up to the first order of $\alpha'$
\begin{eqnarray}
\phi=\phi_{0}+\alpha'\phi_{1},\label{18}\\
{a}_i^2= a_{i0}^2 (1+2\,\alpha'\xi_i),\label{sf}\\
g_{00}=1+2\alpha'\xi_4,\label{17}
\end{eqnarray} 
and applying  a time redefinition which introduces the new time coordinate $\tau$   as  follows \cite{ batakis2}
\begin{eqnarray}\label{tau}
d\tau= a^{-3}{\rm e}^{-\phi}dt,
\end{eqnarray}
where $a^3=a_1a_2a_3$.
Accordingly, the equations of \eqref{ii} and \eqref{33} recast the following equations in  the zeroth order of $\alpha'$ 
\begin{eqnarray}\label{I1}
{(\ln a_{i0} )}''=0,
\end{eqnarray}
\begin{eqnarray}\label{fio}
\phi_0''+\Lambda a_0^{6}{\rm e}^{2\phi_0}=0,
\end{eqnarray}
where $a_0^3=a_{10}a_{20}a_{30}$. Also, in the first order of $\alpha'$ we get
\begin{eqnarray}\label{I2}
\xi_i''-        {(\ln a_{i0} )}'\xi_4'+\hat{K}_{i}=0,
\end{eqnarray}
\begin{eqnarray}\label{fi1}
\begin{aligned}
\phi_1''+2\,\Lambda a_0^{6}{\rm e}^{2\phi_0}(\phi_1&+\sum \xi_i+\xi_4)-\phi_0'\,\xi_4'\\
&-\hat{K}_{\phi}-\sum\hat{K}_{i}-\hat{K}_{0}=0,
\end{aligned}
\end{eqnarray}
where the prime stands for derivations with respect to $\tau$ and the $\hat{K}$ terms are the rewritten versions of $K$ terms in the new time coordinate, multiplied with a $a^{6}{\rm e}^{2\phi}$ factor. Also, the initial value equation \eqref{ini} reads \footnote{We have employed the series expansion of
	$
\ln{a_i}'=\ln{a_{i0}}'+\frac{\alpha'\xi_i'}{1+\alpha'\xi_i}=\ln{a_{i0}}'+\alpha'{\xi_i'}+{\cal{O}} (\alpha'^2)
$}
\begin{eqnarray}\label{initial}
\begin{aligned}
\frac{1}{2}\big[&\sum_{i< j}(\ln a_{i0}^2{\rm e}^{\phi_0})'(\ln a_{j0}^2{\rm e}^{\phi_0})'-\phi_0'^2+2\Lambda\,a_0^6{\rm e}^{2\phi_0}\big]\\
&+\alpha'\big[(\phi_0'+\sum \ln a_{i0}')(\phi_1'+\sum \xi_j)-2\,\sum \ln a_{i0}'\xi_i'\\
&+2\,\Lambda a_0^{6}{\rm e}^{2\phi_0}(\phi_1+\sum \xi_i+\xi_4)+\hat{K}_{\phi}+2\hat{K}_0
\big]=0.
\end{aligned}
\end{eqnarray}

Solution of \eqref{I1} and \eqref{fio} gives the zeroth order of scale factors and dilaton field as follows
\begin{eqnarray}\label{45}
a_{i0}=L_i{\rm e}^{q_i\tau},
\end{eqnarray}
\begin{eqnarray}\label{fioa}
\phi_0=-\sum q_i \tau-\ln (\frac{\sqrt{\Lambda} L_1L_2L_3}{n}\,\cosh (n\tau)),
\end{eqnarray}
where  $L_i, q_i, n$ are integrating constants.
Accordingly, the leading order  string frame Ricci scalar and kinetic of dilaton field are given by
\begin{eqnarray}
R={\frac { \Lambda}{{n}^{2}}}\cosh^2
( n\tau   ) (   (\sum q_i)^2+\sum q_i (q_i
  +2n\tanh  ( n\tau   ) 
)  ),~~
\end{eqnarray} 
\begin{eqnarray}
\begin{aligned}
\dot{\phi_{0}}^2=&a_0^{-6}e^{-2\phi_0}\phi'^2=\frac{\Lambda}{n^2} ((n^2+(\sum q_i)^2)\cosh^2(n\tau)\\
&+2n\,\sum q_i \cosh(n\tau)\sinh(n\tau)-n^2),
\end{aligned}
\end{eqnarray}
which are  growing function of time. The dependence of $R$ and ${\dot{\phi_{0}^2}}$  on $\Lambda$, which implies that the curvature and dilaton field kinetic 
are  comparable with inverse of $\alpha'$ \footnote{Especially,
	if $ ( (\sum q_i)^2+\sum q_i^2)>\frac{n^2}{2 }$, the Ricci is  in $R\alpha'\gtrsim1$ limit even  in $\tau=0$.},  demonstrates the necessity of including the higher orders of $\alpha'$-corrections. {In this work, we focus on studding the effects of first order of $\alpha'$-corrections.}

Demanding only the first order $\alpha'$-corrections in the solutions of metric, dilaton field and $H$-field, only the  zeroth order   $\alpha'$  terms of $\hat{K}$  will be considered. Hence, substituting the  \eqref{45} and\eqref{fioa} into the rewritten versions of  \eqref{kii}  in $\tau$ coordinate gives the explicit forms of $\hat{K}$.
Now, the solutions of \eqref{I2} and \eqref{fi1} give the general forms of $\alpha'$-corrections of scale factor and dilaton field as follows
\begin{eqnarray}\label{46}
\xi_i=-\iint \hat{K}_{i}\,d\tau\,d\tau+{q_i}
\int\xi_4 \,d\tau+l_i\tau+r_i,
\end{eqnarray}
\begin{eqnarray}\label{fi1s}
\begin{aligned}
\phi_1=& Q_1\,\tanh(n\tau)+Q_2\,(n\tau \tanh(n\tau)-1)+\varphi_0\\
&+\frac{1}{n}(\tanh(n\tau)\int (n\tau \tanh(n\tau)-1)\,g(\tau)\,d\tau\\
&~~~~~-(n\tau \tanh(n\tau)-1)\int \tanh(n\tau)\,g(\tau)\,d\tau)\\
&-(n\tanh(n\tau)+\sum q_i)\int \xi_4 \,d\tau,
\end{aligned}
\end{eqnarray}
in which  $g(\tau)$ is given by
\begin{eqnarray}
g(\tau)=\sum(\frac{-2\,n^2}{\cosh^2(n\tau)}\iint \hat{K}_{i}d\tau d\tau+\hat{K}_{i}) -\hat{K}_{\phi}-\hat{K}_{0}.~~~~
\end{eqnarray}
and the $l_i$, $r_i$,  $\varphi_0$, $Q_1$, and $Q_2$ are  are constants of integration. After some calculations with using Taylor series expansions up to the first order of $\alpha'$, it turns out that these constants have the following roles
\begin{itemize}
	\item $l_i$ corresponds to an infinitesimal change in $q_i$, $q_i\rightarrow q_i+\alpha' l_i$,
	\item $r_i$ is a proper scaling in $x_i$ direction,
	\item $Q_1$ describes  an infinitesimal time displacement, $\tau\rightarrow \tau-\alpha' Q_1$,
\item $Q_2$ corresponds to an infinitesimal change in $n$, $n\rightarrow n\,(1-\alpha' Q_2)$,
\item $\varphi_0$ describes a constant shift in dilaton.
\end{itemize}

Substituting these solutions into the initial value equation \eqref{initial} gives
\begin{eqnarray}\label{ini1}
\begin{aligned}
(&n^2-\sum q_i^2)(1+2\alpha'\xi_4)=2\alpha'\big[n^2Q_2+\frac{1}{2}\hat{K}_{\phi}+\hat{K}_0\\
&+\sum(\frac{n^2}{\cosh^2(n\tau)}\iint  \hat{K}_i\,d\tau\,d\tau-(n\,\tanh(n\tau)\\
&+q_i)\int  \hat{K}_i\,d\tau)+n\int \tanh(n\tau)g(\tau)\,d\tau
\big].
\end{aligned}
\end{eqnarray}
Actually, the $n^2-\sum q_i^2$ term is the initial condition on the constants, which appears in the  one-loop $\beta$-function solutions. However, here
we are not allowed to set it to  zero because the right side of this equation does not vanish.
{A comparison between two sides of this equation proposes   the following initial condition on arbitrary constants}
\begin{eqnarray}\label{I1cons}
\begin{aligned}
n^2(1-2\,\alpha'\,Q_2)-\sum q_i^2=0,
\end{aligned}
\end{eqnarray}
and the remaining terms in the right hand of \eqref{ini1}  fixes the correction of laps function, $\xi_4$. { In  the $\alpha'\rightarrow 0$ limit, the zeroth order initial condition  can be recovered  from \eqref{I1cons}. Arising the $\alpha'\,Q_2$ term in the constraint would not be disappointing because, as we have mentioned before,  $Q_2$ can be related to an infinitesimal change in $n$ which acts as $n^2\rightarrow n^2(1-2\alpha' Q_2)$ up to first order of $\alpha'$. Noting the relation between  $\beta$-function equations and Einstein equations, this initial condition may be regarded as a Hamiltonian constraint which  has been corrected in first order $\alpha'$.}

The explicit forms of $\alpha'$-corrections of metric and dilaton after calculating the integrals in \eqref{46}, \eqref{fi1s} and \eqref{ini1} are presented in appendix \ref{app1}.

\subsection{Solution with non-vanishing $H$-field of a spatially homogeneous (time-dependent) $B$-field }\label{Iclass}

As mentioned before, the forms of $H$-field  can be classified based on the orientation of  its dual,   $H^*$, with respect to the 3D hypersurface of homogeneity  $\Sigma^3$ sections. Accordingly, the three classes of $\rightarrow$, $\uparrow$ and $\nearrow$ denoting the   spatial, time, and time-spatial orientations of $H^*$ have been introduced, respectively, by \cite{batakis3}
$$
H^*=H^*_i(t) dx^i,~ H^*=H^*_0 dt,~ H^*=H^*_0(t) dt+H^*_i(t) dx^i.
$$

With  $\Lambda=0$, the solution of one-loop $\beta$-function equations  have been investigated on Bianchi type  models for $\uparrow$, $\rightarrow$ and $\nearrow$ in classes, respectively in \cite{Batakis1,batakis2,batakis3}.

With a non-vanishing $\Lambda$, we have found
no explicit solution for 
the leading order of $\beta$-functions equations  in the $\uparrow$ and $\nearrow$ classes. Therefore, we keep going with  the $\rightarrow$ class 
with the metric
\eqref{metric} in such a way that, considering a $B$-field which is  a function of time, the $H_{0ij}$  components of $H$-field are allowed to be non-zero  \cite{PhysRevD.51.1569}. On the other hand,   the leading order $\beta$-function equations with metric \eqref{metric}
make  the off-diagonal components of 
$H_{\mu\nu}^2$  vanish. This means that only one of the $H_{0ij}$ may be non-zero. Here, there is no preferred direction and we consider
the
following 3-form of field strength $H$ for simplicity \cite{Batakis1}
\begin{eqnarray}\label{H}
H=\frac{1}{3}{A}(t)(a_1a_2)^2\, (d\,t\wedge\,d\,x^1\wedge d\,x^2),
\end{eqnarray}
Then, the $\beta$-function equations \eqref{betaGR}-\eqref{betafR}, using \eqref{k1}, recast the following forms
\begin{eqnarray}\label{11b}
\begin{aligned}
\dot{H_{k}}+ H_j({\sum}H_{i}& +  \dot{\phi}-H_0)+\frac{1}{2}A^2(a_1a_2)^2\\
&+\alpha'(K_{j}+V_{j})g^{00}=0
,~~~~~j=1,2,
\end{aligned}
\end{eqnarray}
\begin{eqnarray}\label{33b}
\dot{H_{3}}+ H_3({\sum}H_{k} +  \dot{\phi}-H_0)+\alpha'(K_{3}+V_{3})g^{00}=0
,
\end{eqnarray}
\begin{eqnarray}\label{00b}
\begin{aligned}
{\sum}(\dot{H_i}+ H_i^2)+\ddot{\phi}-\frac{1}{2}A^2(a_1&a_2)^2-H_0(\dot{\phi}+{\sum}\dot{H_i})\\
&-\alpha'(K_{0}+V_0)g^{00}=0,
\end{aligned}
\end{eqnarray}
\begin{eqnarray}\label{bgb}
\begin{aligned}
&\dot{A}+{A}(\dot{\phi}+\sum H_i -H_0)+\alpha'g^{00}V_{B}=0,
\end{aligned}
\end{eqnarray}
\begin{eqnarray}\label{fib}
\begin{aligned}
-&2\ddot{\phi}-\dot{\phi}^2-{\sum}(2\,\dot{\phi} H_i+{H_i}^2+2\,\dot{H_i})
- ({\sum} H_i)^2-\Lambda g_{00}\\
&-\frac{1}{2}A^2(a_1a_2)^2-2(\dot{\phi}+\sum H_i)H_0
-\alpha'(K_{\phi}+V_{\phi})g^{00}=0,
\end{aligned}
\end{eqnarray}
where the $K$ terms are the same as given in \eqref{kii} and the $V$ terms, which stand for the $H$-field dependent terms, are given as follows
\begin{widetext}
\begin{eqnarray}\label{V1}
\begin{aligned}
V_{1}=&[\frac{1}{2} ((3 H_2^2+(-H_0+4 H_1) H_2+2 H_0^2+2 H_1^2+2 H_3^2+\dot{H_2}) f+2 H_2^2+(-H_0+4 H_1) H_2+H_0^2-2 H_0 H_1\\
&+3 H_1^2+H_3^2+\dot{H_2}) A^2-((H_0-H_1-H_2) f+\frac{1}{2} H_0-\frac{1}{2} H_2) \dot{A} A+\frac{1}{2}(1+f) \dot{A}^2 ] (a_1a_2)^2+\frac{3}{16} (a_1a_2)^4 A^4,
\end{aligned}
\end{eqnarray}
\begin{eqnarray}\label{V1}
\begin{aligned}
V_{2}=V_1(1\leftrightarrow 2),
\end{aligned}
\end{eqnarray}
\begin{eqnarray}
\begin{aligned}
V_{3}=&-\frac{1}{2}[A \dot{H_3} f+H_3 A((-H_0+H_1+H_2-H_3) f-H_0+H_1+H_2-2 H_3)+H_3 \dot{A}] \,A \,(a_1a_2)^2,
\end{aligned}
\end{eqnarray}
\begin{eqnarray}
\begin{aligned}
V_{0}=\frac{1}{2}[&((2 H_0^2-(H_1+H_2) H_0+3 H_1^2+5 H_1 H_2+3 H_2^2+\dot{H_2}+\dot{H_1}) f-H_0^2+2\,( H_1+ H_2) H_0+3 H_1^2+H_1 H_2+3 H_2^2\\
&+2 (\dot{H_2}+ \dot{H_1})+\dot{H_0}) A^2-\dot{A} (2(H_0-H_1-H_2) f-{3} H_0+H_1+H_2) A+\dot{A}^2 f-A\ddot{A}]\,(a_1a_2)^2+\frac{3}{8} (a_1a_2)^4 A^4,
\end{aligned}
\end{eqnarray}
\begin{eqnarray}
\begin{aligned}
{V}_{B}^{(f=1)}=\frac{1}{2} [&(H_1^2+H_2^2-H_0 (H_1+H_2)+\dot{H_1}+\dot{H_2}) \dot{A}+ A \big((H_1+H_2-H_0) (\dot{H_1}+\dot{H_1})+(H_1+H_2) \sum H_i^2\\
&- H_0 (2(H_1^2+H_2^2+H_1H_2)-H_0 (H_1+H_2)) \big)
- A^2 (15\,\dot{A}+A (13\,(H_1 +H_2)-16\,H_0))(a_1a_2)^2],
\end{aligned}
\end{eqnarray}
\begin{eqnarray}\label{br2}
\begin{aligned}
{V}_{B}^{(f=-1)}=\frac{1}{2} [&(H_1^2+H_2^2-H_0 (H_1+H_2)+\dot{H_1}+\dot{H_2}) \dot{A}+ A \big((H_1+H_2-H_0) (\dot{H_1}+\dot{H_2})+(H_1+H_2) \sum H_i^2\\
&-2 H_0 (H_1^2+H_2^2+H_1H_2-H_0 (H_1+H_2)/2) \big)
-11 A^2 (\dot{A}+A (H_1 +H_2-H_0))(a_1a_2)^2],
\end{aligned}
\end{eqnarray}
\begin{eqnarray}\label{Vfi}
\begin{aligned}
V_{\phi}=&-\frac{1}{2}[\dot{H_1}+\dot{H_2}+H_1^2+(H_2-H_0) H_1+H_2^2-H_0 H_2] A^2(a_1a_2)^2+\frac{1}{4} (a_1a_2)^4 A^4.
\end{aligned}
\end{eqnarray}
Similar to what we have done  to obtain the equation \eqref{33},  adding \eqref{11b}-\eqref{00b}  to  \eqref{fib} gives the following equation
\begin{eqnarray}\label{8}
-\ddot{\phi}-\dot{\phi}(\dot{\phi}+\sum H_i)+\dot{\phi}H_0-\Lambda\,g_{00}+A^2(a_1a_2)^2 +\alpha'(K_{\phi}+V_{\phi}+K_{0}+V_0+\sum (K_i+V_i))g^{00}=0.
\end{eqnarray}
Furthermore,  adding \eqref{8} to  \eqref{00b} and subtracting the \eqref{11b}-\eqref{33b} from it give the initial value equation as follows
\begin{eqnarray}\label{inib}
\dot{\phi}(\dot{\phi}+2\sum H_i)+(\sum H_i)^2-\sum H_i^2+\Lambda\,g_{00}-\frac{1}{2}A^2(a_1a_2)^2 -\alpha'(K_{\phi}+V_{\phi}+2(K_0+V_0))g^{00}=0.
\end{eqnarray}
\end{widetext}
Now, we are going to solve the equations \eqref{11b}-\eqref{33b}, \eqref{bgb} and \eqref{8} subject to the initial value equation \eqref{inib}. Again, the equations will be rewritten in the new time coordinate \eqref{tau}  with applying the given series expansion in $\alpha'$ \eqref{18}-\eqref{17}. Also, we will conveniently set
\begin{eqnarray}A(t)=\dot{\eta}=\eta'a^{-3}{\rm e}^{-\phi},
\end{eqnarray} 
and  take the $\alpha'$ expansion  of $\eta$ as
\begin{eqnarray}\label{etaa}
\eta=\eta_0+\alpha'\eta_1.
\end{eqnarray}
Then, in  $\tau$ coordinate, the equations of \eqref{11b}-\eqref{33b}, \eqref{bgb}, and \eqref{8}  lead to the following equations in the zeroth order of $\alpha'$
\begin{eqnarray}\label{23}
{(\ln a_{j0} )}''+\frac{1}{2}\eta_0'^2(a_{10}a_{20})^2=0,\quad j=1,2,
\end{eqnarray}
\begin{eqnarray}
{(\ln a_{30} )}''=0,
\end{eqnarray}
\begin{eqnarray}\label{27}
\eta_0''=0,
\end{eqnarray}
\begin{eqnarray}\label{24}
\phi_0''-\eta_0'^2(a_{10}a_{20})^2+\Lambda a_0^{6}{\rm e}^{2\phi_0}=0.
\end{eqnarray}
The solutions of above equations are found as follows
\begin{eqnarray}\label{ai0clas}
a_{10}=\frac{\sqrt{n}\,{\rm e}^{-q\,\tau}}{\sqrt{b\,L_2 \cosh{(n\tau)}}},
\,
a_{20}=L_2{\rm e}^{2\,q\tau}\,a_{10},\,
a_{30}=L_3{\rm e}^{p\,\tau},\,~~~
\end{eqnarray}
\begin{eqnarray}\label{eta}
\eta_0=b\,\tau,
\end{eqnarray}
\begin{eqnarray}\label{fi0b}
\phi_0=-p\,\tau+\ln{\cosh(n\tau)}-\ln{(\frac{ n\,\Lambda\, L_3}{m\, b}\cosh(m\tau))},
\end{eqnarray}
where $q$, $L_2$, $L_3$, $b$, $p$, $m$,  and $n$ are constant. Obviously, the  $H$-field in this class  brings about an inevitable  anisotropy in the solutions. Compared to the leading order solutions in the absence of $\Lambda$ given in \cite{Batakis1}, only the dilaton field has been modified by the third term in \eqref{fi0b}.
Based on these solutions, the leading order string frame Rici scalar and  kinetic terms   of dilaton field and $B$-field, with $m=n$ for example, are given by
\begin{eqnarray}
R=\frac{\Lambda}{2\,n^2}((-n^2+4\,p^2+4\,q^2)\cosh^2 ( n\tau )-3\,n^2),
\end{eqnarray}
\begin{eqnarray}
\dot{\phi_{0}}^2=a_0^{-6}e^{-2\phi_0}\phi'^2=\frac{p^2\Lambda}{n^2}\cosh^2 ( n\tau ),
\end{eqnarray}
\begin{eqnarray}
H_{\mu\nu\rho}H^{\mu\nu\rho}=6\,\eta'^2\,a_{03}^{-2}e^{-2\phi_0}=6\,\Lambda,
\end{eqnarray} 
Evidently, the string frame Ricci scalar is increasing and starts from the high curvature limit  $R\alpha'\gtrsim1$. Also, the kinetic terms are comparable with the inverse of $\alpha'$. Note that the $R$ and $\dot{\phi_{0}}^2$ keep growing and may dominate the dynamical effect of the $H$-field at late $\tau$.  
The high curvature and  kinetic terms point out the necessity of  considering the $\alpha'$-corrections. { We include the first order $\alpha'$-corrections, noting that the solutions are valid  as long  as the string coupling is weak.  }

In the first order of $\alpha'$ by employing the zeroth order equations \eqref{23}-\eqref{24}, the  equations of \eqref{11b}-\eqref{33b}, \eqref{bgb} and \eqref{8}  read
\begin{eqnarray}\label{25}
\begin{aligned}
\xi_j''-       & {(\ln a_{j0} )}'\xi_4'+(\eta_0'^2(\xi_1+\xi_2)+\eta_0\eta_1)(a_{10}a_{20})^2\\
&+\hat{K}_{j}+\hat{V}_{j}=0,\quad j=1,2,
\end{aligned}
\end{eqnarray}
\begin{eqnarray}
\xi_3''-        {(\ln a_{30} )}'\xi_4'+\hat{K}_{3}+\hat{V}_{3}=0,
\end{eqnarray}
\begin{eqnarray}
\eta_1''-       {\eta_0'}\xi_4'+\hat{V}_{B}=0,
\end{eqnarray}
\begin{eqnarray}\label{26}
\begin{aligned}
\phi_1''+&2\,\Lambda a_0^{6}{\rm e}^{2\phi_0}(\phi_1+\sum \xi_i+\xi_4)\\
&+2\,(\eta_0'^2(\xi_1+\xi_2)+\eta_0\eta_1)(a_{10}a_{20})^2-\phi_0'\,\xi_4'+\rho=0,
\end{aligned}
\end{eqnarray}
where  the $\hat{K}$ and $\hat{V}$ terms are the corresponding terms of \eqref{kii} and \eqref{V1}-\eqref{Vfi}  rewritten in the new $\tau$ coordinate, multiplied with $a^6e^{2\phi}$ factor. Also, the $\rho$ term in \eqref{26} has been defined as
\begin{eqnarray}
\rho=-(\hat{K}_{\phi}+\hat{V}_{\phi}+\sum(\hat{K}_{i}+\hat{V}_{i})+\hat{K}_{0}+\hat{V}_{0}).
\end{eqnarray}
In the same way, the initial value equation \eqref{inib} recasts the following form 
\begin{eqnarray}\label{initialb}
\begin{aligned}
\frac{1}{2}\big[&\sum_{i< j}(\ln a_{i0}^2{\rm e}^{\phi_0})'(\ln a_{j0}^2{\rm e}^{\phi_0})'-\phi_0'^2+2\Lambda\,a_0^6{\rm e}^{2\phi_0}\\
&-\frac{1}{2}\eta_0'^2(a_{10}a_{20})^2\big]+\alpha'\big[(\phi_0'+\sum \ln a_{i0}')(\phi_1'+\sum \xi_j)\\
&-2\,\sum \ln a_{i0}'\xi_i'+2\,\Lambda a_0^{6}{\rm e}^{2\phi_0}(\phi_1+\sum \xi_i+\xi_4)\\
&-(\eta_0'^2(\xi_1+\xi_2)+\eta_0\eta_1)(a_{10}a_{20})^2\\
&+\hat{K}_{\phi}+\hat{V}_{\phi}+2(\hat{K}_0+\hat{V}_0)
\big]=0.
\end{aligned}
\end{eqnarray}
Again, because we are interested in  the first order $\alpha'$-corrections in the solutions of metric, dilaton field and $H$-field,  only the  zeroth order   $\alpha'$  terms of $\hat{K}$ and $\hat{V}$ will be considered, which depend on  $a_{i0}$, $\phi_0$ and $\eta_0$. Hence, substituting the  \eqref{ai0clas}-\eqref{fi0b} into the rewritten versions of  \eqref{kii} and \eqref{V1}-\eqref{Vfi} in $\tau$ coordinate gives the explicit forms of $\hat{K}$ and $\hat{V}$. Then, solving the  equations of \eqref{25}-\eqref{26}  gives the general forms of  $\alpha'$-corrections of scale factors $\xi_i$, laps function $\xi_4$, $H$-field $\eta_1$, and dilaton field $\phi_1$  as follows
\begin{eqnarray}\label{ai1}
\begin{aligned}
\xi_1=& c_1\,\tanh(n\tau)+c_2\,(n\tau \tanh(n\tau)-1)+r_1\\
&+(\frac{n}{2}\tanh(n\tau)- q)\int \xi_4 \,d\tau\\
&+\frac{1}{n}[\tanh(n\tau)\int (n\tau \tanh(n\tau)-1)\,g_1(\tau)\,d\tau\\
&-(n\tau \tanh(n\tau)-1)\int \tanh(n\tau)\,g_1(\tau)\,d\tau],
\end{aligned}
\end{eqnarray}
\begin{eqnarray}
\begin{aligned}
\xi_2=& \xi_1-\iint(\hat{K}_{2}+\hat{V}_{2}-\hat{K}_{1}-\hat{V}_{1})\,d\tau d\tau+2 q\int \xi_4 d\tau,
\end{aligned}
\end{eqnarray}
\begin{eqnarray}\label{ai3}
\begin{aligned}
\xi_3=& -\iint(\hat{K}_{3}+\hat{V}_{3})\,d\tau\,d\tau+p\int \xi_4 \,d\tau+r_3,
\end{aligned}
\end{eqnarray}
\begin{eqnarray}
\begin{aligned}
\eta_1=& -\iint\hat{V}_{B}\,d\tau\,d\tau+b\int \xi_4 \,d\tau,
\end{aligned}
\end{eqnarray}
\begin{eqnarray}\label{fic}
\begin{aligned}
\phi_1=
&\frac{1}{m}[\tanh(m\tau)\int (m\tau \tanh(m\tau)-1)\,g_{\phi}(\tau)\,d\tau\\
&-(m\tau \tanh(m\tau)-1)\int \tanh(m\tau)\,g_{\phi}(\tau)\,d\tau]+\varphi_0,
\end{aligned}
\end{eqnarray}
where   $g_1$, and $g_{\phi}$ are given by
\begin{eqnarray}
\begin{aligned}
g_{1}(\tau)=&-\frac{2n^2}{\cosh^2 (n\tau)}(\iint (\hat{K}_{2}+\hat{V}_{2}-\hat{K}_{1}-\hat{V}_{1})\,d\tau\,d\tau\\
&-q_0^{-1}\int \hat{V}_{B}\,d\tau)+\hat{K}_{1}+\hat{V}_{1},
\end{aligned}
\end{eqnarray}
\begin{eqnarray}
\begin{aligned}
g_{\phi}(\tau)=&2\,\Lambda a_0^{6}{\rm e}^{2\phi_0}(\sum \xi_i+\xi_4)-\phi_0'\,\xi_4'+\rho\\
&+2\,(\eta_0'^2(\xi_1+\xi_2)+\eta_0\eta_1)(a_{10}a_{20})^2,
\end{aligned}
\end{eqnarray}
and  $c_1, c_2, r_i,$ and $\varphi_0$ are integrating constants. A closer look with using Taylor series expansion reveals that up to first order of $\alpha'$ 
\begin{itemize}
	\item $c_1$  corresponds to an infinitesimal time displacement $\tau\rightarrow\tau-2\alpha'c_1$,
\item $c_2$ acts as an infinitesimal change in n, $n\rightarrow n(1-2\alpha'c_2)$,
\item $\varphi_0$ is an infinitesimal shift in dilaton,
\item $r_i$ is a proper scaling in $x_i$ direction.
\end{itemize}

 For calculating the integrals of \eqref{fic} it is convenient to set $m=n$, where as given in appendix \ref{App},  the $\phi_1$ and $g_{\phi}$ take the forms of \eqref{fic1} and \eqref{gfi}.  Then, similar to what has been done in the previous subsection,  substituting the above solutions into the initial value equation \eqref{initialb} gives
\begin{eqnarray}\label{91}
\begin{aligned}
&\frac{1}{2}
(n^2-2\,p^2-4\,q_1^2)(1+2\alpha'\,\xi_4)=\alpha'[2\,n^2c_2\\
&-2n\tanh(n\tau)\int(\hat{K}_{3}+\hat{V}_{3})\,d\tau+\hat{K}_{\phi}+\hat{V}_{\phi}+2\hat{K}_{0}+2\hat{V}_{0}]\\
&+(n\tanh(n\tau)+2\,q_1)\int(\hat{K}_{2}+\hat{V}_{2}-\hat{K}_{1}-\hat{V}_{1})\,d\tau\\
&-\frac{n^2}{\cosh^2 (n\tau)}(q_0^{-1}\int\hat{V}_{B}\,d\tau+\iint(\hat{K}_{2}+\hat{V}_{2}-\hat{K}_{1}-\hat{V}_{1}\\
&-2\hat{K}_{3}-2\hat{V}_{3})\,d\tau\,d\tau)-2n\int \tanh(n\tau)\,(g_{\phi}(\tau)+g_1(\tau))\,d\tau.
\end{aligned}
\end{eqnarray}
Again, we have a similar situation as discussed in the absence of $H$-field case in \eqref{ini1} and this equation leads to the following condition on the arbitrary constants
\begin{eqnarray}\label{ini2}
\begin{aligned}
n^2(1-4\,\alpha'c_2)-2\,p^2-4\,q^2=0.
\end{aligned}
\end{eqnarray}
In  $\alpha'\rightarrow 0$ limit, the initial condition of $n^2-2\,p^2-4\,q^2=0$ which appears in the solutions of one-loop $\beta$-functions can be recovered. 
But in two-loop order, this condition has been modified by a term in the order $\alpha'$ and   $n^2-2\,p^2-4\,q^2\neq 0$ is required for consistency in the solution of \eqref{91}.
Then, the remaining terms in \eqref{91} fix the correction of lapse function, $\xi_4$.

Now,  calculating the integrals  in \eqref{ai1}-\eqref{fic}  give the explicit forms of first $\alpha'$-correction  of scale factors, dilaton field and $H$-field. Because of dense mathematical results, the  final forms of $\xi_i$, $\xi_4$, $\phi_1$ and $\eta_1$ in two RS of Gauss-Bonnet and $R^2$ are presented in Appendix  \ref{App}. 
\section{The Einstein frame representation}\label{sec4}
Having solved the two-loop $\beta$-function equations  in two cases of  vanishing $H$-field  and presence of a time-dependent $H$-field in section \ref{sec3}, we return to Einstein frame  field equations \eqref{22} in order to study the cosmological implications  of the $\alpha'$-corrected solutions.
{ As mentioned in section \ref{sec2}, in the   $\alpha'$ order of effective actions there is a field redefinition ambiguity and a class of equivalent effective actions corresponding to the same $S$-matrix. The cosmological effects of the field redefinition have been studied with constant dilaton in \cite{akune2006inflation}, and with a time-dependent dilaton in \cite{maeda2012accelerating}, where a generalized effective action obtained by the field redefinitions has been investigated. However, here we consider  the Gauss-Bonnet effective action \eqref{GBaction}. } In Einstein frame, regarding $T_{\mu}^{\mu}=(-\rho,P_1,P_2,P_3)$, the
non-zero components of energy-momentum tensors  \eqref{Tfi}-\eqref{TB2} give the  effective  energy density and pressures based on \eqref{Teff}.  The Einstein frame metric, which is related to the string frame metric by the conformal transformation \eqref{3},
is considered as follows
\begin{eqnarray}\label{metrice}
ds^2=\tilde{g}_{\mu\nu}\,dx^{\mu}\,dx^{\nu}=-d\tilde{t}^{2}+ \sum_{i}^3\tilde{a}_i^2\,(dx^{i})^{2},
\end{eqnarray}
where  $\tilde{a}_{i}$ are the Einstein frame scale factors related to the string frame ones by $\tilde{a}_{i}^2=e^{\phi}a_i^2$, in such a way that up to the first order of $\alpha'$ we have
\begin{eqnarray}
\tilde{a}_{i}^2=e^{\phi_0}a_{i0}^2(1+\alpha'(2\,\xi_i+\phi_1)).
\end{eqnarray}
Also, the time element in \eqref{metrice} has been defined using the  following relation
\begin{eqnarray}
\begin{split}
d\,\tilde{t}=e^{\frac{\phi}{2}}\,\sqrt{g_{00}}\,d\,t.
\end{split}
\end{eqnarray}
In the previous section, we found the solutions in terms of the time coordinate $\tau$ which, using \eqref{tau}, has the following relation with the cosmic-time $\tilde{t}$ 
\begin{eqnarray}\label{d}
\begin{split}
d\,\tilde{t}={a_0}^{3}e^{{3}\phi_0/2}(1+\alpha'(\sum \xi_i+\xi_4+\frac{3}{2}\phi_1))d\,\tau.
\end{split}
\end{eqnarray}
The integrating of  above expression and transforming from $\tau$ to $\tilde{t}$  are not straightforward in the solutions; hence, for investigating the behavior of solutions, the time derivatives in  physical quantities in Einstein frame will be rewritten in terms of $\tau$-derivatives such as follows
\begin{eqnarray}\label{adot}
\begin{split}
\dot{\tilde{a}}_i\equiv \frac{d\, \tilde{a}_i}{d\,\tilde{t}}={\tilde{   a}^{-3}}({\tilde{a}_{i0}'+\alpha'(\tilde{a}_{i0}(\xi_i+\frac{1}{2}\phi_1))'}),
\end{split}
\end{eqnarray}
\begin{eqnarray}\label{a2dot}
\begin{aligned}
\ddot{\tilde{a}}_i\equiv\frac{d^2\, \tilde{a}_i}{d\,\tilde{t}^2}=\tilde{        a}^{-6}\big(&\tilde{a}_{i0}''-\tilde{a}_{i0}'\sum \ln{\tilde{a}_{j0}}'+\alpha'[(\tilde{a}_{i0}(\xi_i+\frac{1}{2}\phi_1))''\\
&-\tilde{a}_{i0}'(\ln(1+\sum \xi_j+\xi_4+\frac{3}{2}\phi_1))'\\
&+\alpha'(\tilde{a}_{i0}(\xi_i+\frac{1}{2}\phi_1))'\sum \ln{\tilde{a}_{j0}}']   \big).
\end{aligned}
\end{eqnarray}
Here and hereafter, the dot symbol stands for derivation with respect to $\tilde{        t}$. Also, the second derivative of averaged scale factor up to first order of $\alpha'$ is given by
\begin{eqnarray}\label{av2dot}
\begin{aligned}
\ddot{\tilde{a}}=&\frac{d^2\, }{d\,\tilde{t}^2}(\tilde{a}_1\tilde{a}_2\tilde{a}_3)^{\frac{1}{3}}=\frac{1}{9}(\tilde{a}_1\tilde{a}_2\tilde{a}_3)^{\frac{1}{3}}(3\sum \dot{\tilde{H}}_i-(\sum \tilde{H}_i)^2)\\
=&\tilde{       a}^{-5}\big((\ln\tilde{a}_{i0})''-(\ln\tilde{a}_{i0})'\sum (\ln\tilde{a}_{i0})'+\alpha'[(\xi_i+\frac{1}{2}\phi_1)''\\
&-(\ln\tilde{a}_{i0})'(\ln(1+\sum \xi_j+\xi_4+\frac{3}{2}\phi_1))'\\
&+\alpha'(\xi_i+\frac{1}{2}\phi_1)'\sum \ln{\tilde{a}_{j0}}']   \big).
\end{aligned}
\end{eqnarray}
In the equation \eqref{d},  if the coefficient term of $d\tau$  is positive, $\tilde{t}$ will be an increasing function of $\tau$ and then $d\tau>0$ if and only if $d\tilde{t}>0$. In this sense, the early and late behaviors of solutions can be investigated in $\tau\rightarrow  0$ and $\tau\rightarrow \infty$ limits.

The solutions of $\beta$-function equations for metric, dilaton field and $B$-field contain integrating constants which are allowed to be any real number provided that some of them satisfy the initial conditions of \eqref{I1cons} in vanishing $H$-field case and \eqref{ini2} in the presence of $H$-field.  It turns out that the appeared constants in the zeroth order solutions effect the general behavior of solutions while the constants of first $\alpha'$-corrections influence the early time behavior. As a matter of fact that it is not convenient  to predict the cosmological behavior of solutions without selecting some values for these constants, we are going to investigate the features of obtained solutions with some chosen set of arbitrary constants. In this regard, besides the $\beta$-functions prescribed initial conditions of  \eqref{I1cons}  and \eqref{ini2}, the other conditions   that can be demanded from the cosmological point of view to be imposed
on the obtained solutions are the positive sign of coefficient term of $d\tau$ in \eqref{d}, satisfying the energy condition  $\rho^{\mathrm{(eff)}}>0$ and having no singularity corresponding to the vanishing of scale factors in future. 
Also, it should be noticed that the calculation of $\alpha'$-corrections is trusted  as long as  the string coupling  at tree-level of the string interaction is weak, i.e. $g_s\ll 1$. Hence, the reliable area of solutions may be affected by selected  parameters.

\subsection{The evolution with vanishing $H$-field }\label{H=0e}

In the absence of $H$-field, the solutions of two-loop $\beta$-function equations were found in section \ref{H=0}. In  Einstein frame,
according to \eqref{Tfi}, we have the  dilaton field and charge deficit term $\Lambda$ contributions to the effective energy density and pressures up to first order of $\alpha'$ as follows
\begin{eqnarray}\label{rofi0}
\begin{aligned}
\rho^{(\phi)}&=\frac{1}{4}(\dot{\phi}^2-2\Lambda\,e^{-\phi})\\
&=\frac{1}{4}(\dot{\phi}^2_0-2\Lambda\,e^{-\phi_0}+\alpha'(2\,\dot{\phi_0}\dot{\phi_1}+2\Lambda\,e^{-\phi_0}\phi_1)),
\end{aligned}
\end{eqnarray}
\begin{eqnarray}\label{pfi0}
\begin{aligned}
P_i^{(\phi)}&=\frac{1}{4}(\dot{\phi}^2+2\Lambda\,e^{-\phi})\\
&=\frac{1}{4}(\dot{\phi}^2_0+2\Lambda\,e^{-\phi_0}+\alpha'(2\,\dot{\phi_0}\dot{\phi_1}-2\Lambda\,e^{-\phi_0}\phi_1)).
\end{aligned}
\end{eqnarray}
Also, the $T^{\mathrm{(GB)}}_{\mu\nu}$ \eqref{TGB} gives
\begin{eqnarray}
\rho^{(\mathrm{GB})}=-3\,\alpha'e^{\phi}\dot{\phi}\tilde{H}_1\tilde{H}_2\tilde{H}_3,
\end{eqnarray}
\begin{eqnarray}
\begin{aligned}
P_1^{(\mathrm{GB})}=\alpha'e^{\phi}(&\tilde{H}_2\tilde{H}_3(\ddot{\phi}+\dot{\phi}^2)+\dot{\phi}(\tilde{H}_2\dot{\tilde{H_3}}+\tilde{H}_3\dot{\tilde{H_2}}\\
&+\tilde{H}_2\tilde{H}_3(\tilde{H}_2+\tilde{H}_3)), 
\end{aligned}
\end{eqnarray}
\begin{eqnarray}
\begin{aligned}
P_2^{(\mathrm{GB})}=\alpha'e^{\phi}(&\tilde{H}_1\tilde{H}_3(\ddot{\phi}+\dot{\phi}^2)+\dot{\phi}(\tilde{H}_1\dot{\tilde{H_3}}+\tilde{H}_3\dot{\tilde{H_1}}\\
&+\tilde{H}_1\tilde{H}_3(\tilde{H}_1+\tilde{H}_3)),
\end{aligned}
\end{eqnarray}
\begin{eqnarray}
\begin{aligned}
P_3^{(\mathrm{GB})}=\alpha'e^{\phi}(&\tilde{H}_2\tilde{H}_1(\ddot{\phi}+\dot{\phi}^2)+\dot{\phi}(\tilde{H}_2\dot{\tilde{H_1}}+\tilde{H}_1\dot{\tilde{H_2}}\\
&+\tilde{H}_2\tilde{H}_1(\tilde{H}_2+\tilde{H}_1)),\label{pgb}
\end{aligned}
\end{eqnarray}
where the $\tilde{H}_i$ are Einstein frame Hubble  parameters defined by $\tilde{H}_i=      \frac{d\, }{d\,\tilde{t}}\ln \tilde{a}_i$ \footnote{Up to first order of $\alpha'$ we have $\tilde{H_i}=\frac{d}{d\tilde{   t}}\ln \tilde{a}_{i0}+\alpha'\frac{d}{d\tilde{ t}}(\xi_i+\frac{1}{2}\phi_1)$. }. It should be noted that for investigating  the solutions up to first order of $\alpha'$,  only the zeroth order of $\tilde{H_i}$ and $\phi$ contribute in the $T^{\mathrm{(GB)}}_{\mu\nu}$ and  $T^{\mathrm{(B_2)}}_{\mu\nu}$.

The  $\alpha'$-corrected string frame scale factors, laps function, and dilaton field  introduced in  \eqref{18}-\eqref{17} have been found in the forms of \eqref{45}, \eqref{fioa} and \eqref{xiI}-\eqref{xi4b0} in the absence of $H$-field. 
The leading order  string coupling, in this case, is  given by 
\begin{eqnarray}\label{gsh=0}
g_s={\rm e}^{-\phi_0}=\frac{\sqrt{\Lambda}L_1L_2L_3}{n}\,\cosh(n\tau)\,{\rm e}^{\sum q_i \tau}.
\end{eqnarray}
Aiming at considering only the $\alpha'$-corrections, the string coupling is required to be weak. In early $\tau$ it can be achieved by setting the $L_i$ constants at least of order $\sqrt{\alpha'}$. Then,
with $\sum q_i<0$ 
the weak string coupling condition is satisfied in all times with proper selecting of $n$, but with $\sum q_i>0$ the $g_s$ may leave the weak coupling limit as time goes on. 
Here, we present two examples with isotropic and anisotropic parametrization. The appeared constants in the solutions are $q_i, n, l_i,  L_i, Q_1$, and  $Q_2$ where  $q_i, n$, and  $Q_2$ have to satisfy the initial condition \eqref{I1cons}.

As an isotropic example we set $q_i=-2$, $n=1$, $L_i=\sqrt{\alpha'}$,  $l_i=-10$,  $r_i=0$, $Q_1=500$, along with using the \eqref{I1cons} condition for fixing  $Q_2$. Here, the string frame curvature is increasing which starts in $R\alpha'\gtrsim 1$ limit and the string coupling is weak and decreasing. These imply that the  $\alpha'$-corrected solutions are valid and important even in late times and the  string loop correction can be ignored. 
This example has positive but decreasing $\dot{\tilde{a}}$ and  $\ddot{\tilde{a}}$.
Effectively, the energy density is decreasing and  pressure is negative and increasing.      
Also,  the  strong energy condition is violated by $\rho^{\mathrm{(eff)}}+3\,P^{\mathrm{(eff)}}<0$, where the null energy condition $\rho^{\mathrm{(eff)}}+P^{\mathrm{(eff)}}\geq0$ is satisfied except  in a short range of time near $\tau=0$.
Hence, this example describes an accelerated expanding universe with
avoidance of
initial singularity related to violation of strong energy condition. The phantom phase, which by definition satisfies the $P^{\mathrm{(eff)}}<-\rho^{\mathrm{(eff)}}$  with the equation of state parameter $w$ less than $-1$, is transient in early time.

Having found no compatible example with $\sum q_i<0$ in anisotropic case with preserving the signature of metric \eqref{metrice} and the energy condition of $\rho^{\mathrm{(eff)}}>0$, 
we set $q_1=1.1$, $q_2=1.5$, $q_3=2$, $n=2.4$, $l_i=1$, $r_i=0$, and $L_i=20\sqrt{\alpha'}$. The $Q_2$   is fixed by initial condition \eqref{I1cons}.  
This example gives positive and increasing  $\dot{\tilde{a}}_i$, $\ddot{\tilde{a}}_i$ and $\ddot{\tilde{a}}$  defined in \eqref{adot}, \eqref{a2dot} and \eqref{av2dot}. 
Furthermore, the $P_i^{\mathrm{(eff)}}$ are negative
with violating the strong and null energy conditions. It is worth mentioning that as time goes on in this parametrization,  accompanied by the unbounded growth of curvature which has started in $R\alpha'\gtrsim 1$ limit, the string coupling $g_s$ keeps growing until
leaving the weak coupling limit. Reaching the strongly coupled high curvature phase signals the entering of the system into the full M-theory regime \cite{witten1995string,hovrava1996heterotic}. Nevertheless,  this given solution is valid as long as the $g_s$ is sufficiently weak in early $\tau$  and describes an accelerated expansion in all directions with avoidance of initial singularity and behaves as phantom with $w_i<-1$. Existence of the $w < −1$ region   opens up the possibility of occurring the so-called Big-rip singularity which has been classified in four classes   \cite{leith2007gauss,nojiri2005properties,barrow2004sudden,barrow1986closed}. Noting the scale factors and dilaton field given by \eqref{45}, \eqref{fioa} and \eqref{xiI}-\eqref{xi4b0},  no finite time singularity appears in the scale factors, derivatives of Hubble parameters, dilaton field (and their derivatives) and consequently,  according to \eqref{rofi0}-\eqref{pgb}, in the pressures and energy density. Exhibiting no sudden future time divergence by these quantities, which is also verified by their plots, implies that none of the four types of Big-Rip singularity occurs.

\subsection{The evolution with the non-vanishing $H$-field }\label{Hnon0}

Considering a spatially homogeneous time-dependent  $B$-field with the field strength tensor of type \eqref{H}, the solutions of two-loop $\beta$-function equations   have been found in section \ref{Iclass}. A usual effect of this type of $H$-field  is an anisotropic evolution in spatial directions. Also,    the contribution of $B$-field   brings up
a RS dependence on the two-loop order $\beta$-function equations, and consequently in their solutions. We have considered two special RS of $R^2$ and Gauss-Bonnet corresponding to the RS parameters of $f=1$ and $f=-1$, respectively.
In   Gauss-Bonnet scheme, with the $H$-field of form  \eqref{H}, the components of energy-momentum tensors \eqref{TB} and \eqref{TB2}  recast the following forms
\begin{eqnarray}
\rho^{(B_1)}=-P_1^{(B_1)}=-P_2^{(B_1)}=P_3^{(B_1)}=\frac{1}{4}{A}^2\tilde{a}_1^2\tilde{a}_2^2e^{2\,\phi},~~~~
\end{eqnarray}
\begin{eqnarray}\label{tb1}
\begin{aligned}
\rho^{(B_2)}=&\frac{\alpha'}{64} {A}^2 \tilde{a}_1^2\tilde{a}_2^2e^{-\phi} (\frac{38}{3}{A}^2\tilde{a}_1^2\tilde{a}_2^2 e^{-2 \phi}  +15 \tilde{H}_1\tilde{H}_2\\
&-8\, (\tilde{H}_1+\tilde{H}_2) (\tilde{H}_3+3\,\dot{\phi})+32\, (\tilde{H}_2^2+\tilde{H}_1^2)\\
&+4 \dot{\phi}^2+48\, (\dot{{\tilde{H_1}}}+\dot{\tilde{H_2}})), 
\end{aligned}
\end{eqnarray}
\begin{eqnarray}\label{Bp1}
\begin{aligned}
P_1^{(B_2)}=&\frac{\alpha'}{64}  \tilde{a}_1^2\tilde{a}_2^2e^{-\phi} \big(-\frac{26}{3} {A}^4\tilde{a}_1^2\tilde{a}_2^2 e^{-2 \phi}+8\,\dot{A}^2+(72 \dot{\phi}^2\\
&-8 (6 \tilde{H}_1-\tilde{H}_2) \dot{\phi}-16 \tilde{H}_1^2- (31 \tilde{H}_2+32 \tilde{H}_3) \tilde{H}_1\\
&-80 \tilde{H}_2^2-8\, \ddot{\phi}-24 \tilde{H}_2\tilde{H}_3-64 \dot{\tilde{H}}_1-48  \dot{\tilde{H}}_2\\
&-24  \dot{\tilde{H}}_3) {A}^2+16\,(3 \tilde{H}_1 \dot{A}+2\tilde{H}_2  \dot{A}-3 \dot{\phi}  \dot{A}+ \ddot{A})  {A}\big), 
\end{aligned}
\end{eqnarray}
\begin{eqnarray}
\begin{aligned}
P_2^{(B_2)}=P_1^{(B_2)}(1 \leftrightarrow 2)
,
\end{aligned}
\end{eqnarray}
\begin{eqnarray}
\begin{aligned}
P_3^{(B_2)}=\frac{-\alpha'}{192} {A}^2 \tilde{a}_1^2\tilde{a}_2^2e^{-\phi} \big(&10 {A}^2\tilde{a}_1^2\tilde{a}_2^2 e^{-2 \phi}+36\,\dot{\phi}^2\\
&-3\tilde{H}_1\tilde{H}_2\big). 
\end{aligned}
\end{eqnarray}

The  contributions of dilaton field and Gauss-Bonnet term  in the effective energy-momentum tensor are the same as given  by \eqref{rofi0}-\eqref{pgb}. Again, only the zeroth order of $\tilde{a}_i$, $\tilde{H}_i$, $A$ and $\phi$ will be effective in the components of $T_{\mu\nu}^{(GB)}$ and $T_{\mu\nu}^{(B_2)}$.
The $H$-field in this class brings about anisotropic pressures which means that it acts like an anisotropic fluid. Its contribution to the pressures in the zeroth order of $\alpha'$, given by  \eqref{tb1},  is negative in the $x^1$ and $x^2$ directions and positive in $x^3$ direction; but the signs  in the first order of  $\alpha'$ may be affected by selected values for the arbitrary constants of solutions.

In section \ref{Iclass}, the solutions of two-loop $\beta$-function equations gave the $\alpha'$-corrected string frame scale factors, laps function, dilaton field, and $H$-field  of  \eqref{18}-\eqref{17} and \eqref{etaa} in the forms of \eqref{ai0clas}-\eqref{fi0b} and \eqref{ai1}-\eqref{fic}, where the final forms of correction terms after performing the integrals are presented in appendix \ref{App} through \eqref{aicals1}-\eqref{etar2}.
Regarding the obtained scale factors, it turns out that the anisotropy is inevitable with the $H$-field in this considered class.
Here, we are going to investigate the behavior of solutions by choosing some values for arbitrary constants in two  RS and study the feature of energy-momentum tensor components of Gauss-Bonnet scheme. The appeared constants in the solutions are $q, p, n, b,  c_1, c_2, r_i,  $ and $ L_3 $ where the $p, q, n, $ and $c_2$ must satisfy the initial condition \eqref{ini2}. 

According to \eqref{fi0b}, the leading order string coupling is given by
\begin{eqnarray}\label{gsclas}
g_s=e^{-\phi_0}={L_3}{b}^{-1}\sqrt{\Lambda}\,e^{p\tau}.
\end{eqnarray}
Its value  in the origin of $\tau$ can be set to be sufficiently small, for instance by letting the $L_3$ and $b^{-1}$ constants to be in order $\sqrt{\alpha'}$. The behavior of $g_s$ depends on the sign of $p$, in such a way that with $p\leq 0$ the weak coupling condition is always satisfied, but with 
$p>0$ it increases going forward in time and may leave the weak coupling limit.

In  the $R^2$ scheme, as an example  the set of    $q = 3, p = -2, n = 3, L_3 = 2\sqrt{\alpha'}, b^{-1} = 2\sqrt{\alpha'}, \varphi_0=12,    c_1 = 1,    r_i   
  = 0$ can be chosen. Investigating    $\dot{\tilde{a}}_i$ shows that this example is expanding in all direction  in early times, then turns to  Kasner-type expanding, i.e. expanding in two directions and contracting in one direction with  $\dot{\tilde{a}}_1<0$, followed by an  expanding  in all directions phase. Moreover, the behavior $\ddot{\tilde{a}}_i$ are as follows: $\ddot{\tilde{a}}_1$ is negative in early times and then turns to be positive, the $\ddot{\tilde{a}}_2$ is  positive forever and the $\ddot{\tilde{a}}_3$ has similar behavior  to the first direction but leaves the negative phase earlier \footnote{Setting $q=-3$ just exchanges the behavior of first and second direction.}. Also, the first and second derivative of averaged scale factor \eqref{av2dot}
is positive which shows that the expansion is accelerated.

In the  Gauss-Bonnet scheme, for example  the parametrization of $ q = 1, p = 3.15, n = 5, L_3 = \sqrt{\alpha'}, b^{-1} = \sqrt{\alpha'},    \varphi_0 = 300, r_3=350, c_1= r_2= r_2 = 0$, with using  \eqref{initialb} to  fix $c_2$, is capable of preserving  $\rho^{\mathrm{(eff)}}>0$, preventing the vanishing of scale factors and making the coefficient of $d\tau$ in \eqref{d} to be positive. This example starts   expanding in all directions  and then becomes contracting in first and third directions, where the $\dot{\tilde{a}}_1$ leaves the negative area earlier than  $\dot{\tilde{a}}_3$,  and then becomes FRW type expansion along all directions. In addition, investigating the behavior of  $\ddot{\tilde{a}}_i$  \eqref{a2dot} shows that $\ddot{\tilde{a}}_1$ and $\ddot{\tilde{a}}_2$  are  negative at first and then become positive and keep increasing, but  $\ddot{\tilde{a}}_3$ is negative and decreasing. Also, the averaged scale factor has $\ddot{\tilde{a}}>0$,
which implies that the evolution of model is accelerated.
Furthermore, $P_1^{\mathrm{(eff)}}$ and $P_2^{\mathrm{(eff)}}$  start negatively and keep increasing  to become  positive, where   $P_3^{\mathrm{(eff)}}$   is negative and decreasing.  
Effectively, the strong energy condition $\rho^{\mathrm{(eff)}}+\sum P_i^{\mathrm{(eff)}}>0$  is violated   and hence the initial singularity is avoided.
Also, the null energy condition is violated in the third direction so the   time-dependent equation of state parameter in the third direction is $w_3<-1$.  However,  investigating the $\tilde{a}_i$, $\dot{\tilde{H_i}}$, $\rho^{\mathrm{(eff)}}$ and $P^{\mathrm{(eff)}}$ shows that there is no evidence of Big-Rip singularity occurrence corresponding to a sudden divergence in these quantities. It is worth mentioning that  the $g_s$,  starting in the weak coupling limit in early time, evolves toward the strong coupling because the  $p$ is
positive here. Hence, as a matter of fact that curvature and $g_s$ show unbounded growth, the calculation of $\alpha'$-corrections  is
no longer valid  in $\tau>p^{-1}\ln(\frac{b}{L_3\Lambda})$,  when the condition of $g_s\ll 1$ is violated and universe  enters  the non-perturbative regime of the M-theory.

\section{Conclusion}\label{Conclusion}
The higher-derivative corrections are introduced to the string effective action when the equivalence between  field equations and higher loop $\sigma$-model $\beta$-functions is considered. Aimed at presenting non-critical Bianchi type $I$ string cosmology solutions, we have solved the  $\beta$-function equations in the presence of  central charge deficit term $\Lambda$. Being of order $\alpha'^{-1}$, the $\Lambda$ term resulted in the leading order string curvature in the high curvature limit of $R\alpha'\gtrsim1$, which requires the considering of  higher order   $\beta$-function equations and consequently including the $\alpha'$-corrections in the effective action.
 The other type  of effective action modification, i.e. the stringy loop corrections,  have been assumed to be negligible which is reliable where the leading order sting coupling is  weak, i.e. $g_s\ll 1$.

Considering the two-loop (order $\alpha'$) $\beta$-function equations with $\Lambda\neq 0$ in two cases of vanishing and non-vanishing  $H$-field, we have calculated their solutions implementing a perturbation series expansion up to first order of $\alpha'$ on the background fields. The solutions provided  $\alpha'$-corrected string frame metric, dilaton field and $H$-field. Then, in order to study the cosmological implications of solutions,  the corresponding solutions in Einstein frame have been obtained by performing a conformal transformation on the metric.
Also, the  Einstein frame effective action has been considered to include the Gauss-Bonnet term coupled to the dilaton field, because from string theory point of view the Gauss-Bonnet combination is indistinguishable from the other quadratic curvature corrections. In this sense, the effective  energy-momentum tensor in the Einstein frame field equations contains the contributions of 
Gauss-Bonnet term, dilaton, $\Lambda$, and $H$-field.

For investigating the detailed behaviors of the   $\alpha'$-corrected background field solutions and the effective energy density and pressures, we have considered some set of values for the appeared arbitrary constants in the solutions. These constants are allowed to be any real number provided that some of them satisfy the initial condition. Particularly, preserving the energy condition  $\rho^{\mathrm{(eff)}}>0$ and avoiding the singularity in metric caused by vanishing scale factors have been demanded in selecting the constants. 
Without the contribution of $H$-field the solutions are not necessarily anisotropic and two examples with choosing isotropic and anisotropic parametrization were discussed. In the isotropic case, an example describing accelerated expanding universe with a transient phantom phase in early time was presented. In the offered
anisotropic example,  dilaton field starts from weak coupling in early time and evolves to strong coupling regime as time goes on. Hence, the calculation of $\alpha'$-corrections is valid only in  sufficiently small times with $g_s\ll 1$,
where the given example describes an accelerated expansion which crosses the phantom phase $w_i<-1$  with violating the null energy condition in all directions.
In addition, in the presence of time-dependent $H$-field whose $H_{012}$ component was considered to be non-zero,  the solutions appeared to be inevitably anisotropic. In this case, the presented example with the chosen set of constants describes an accelerating model evolving from a   Kasner-type phase to  FRW-type expansion in all directions along with violating the null energy condition in the third direction. Its valid cosmological era is limited to early times by growing of the $g_s$.

It is worth mentioning that  the conformal invariance condition prescribes including the whole $\alpha'$-correction series. { Especially, working at high curvature limit, all higher orders of $\alpha'$-corrections certainly become important.  Nevertheless, aimed at finding a pattern given by including the corrections, we restricted our discussion to the first order $\alpha'$ (two-loop) $\beta$-function equations as the solutions of  first order $\alpha'$-corrected string effective action at the zeroth order in the string coupling.} Even in the first order, the corrections have been capable of i) excluding the initial singularity in the regime of violation of strong energy condition, and ii) describing the accelerated expansion of the universe. 
However, as time passes, the validity of examples may be restricted by growing of the string coupling and passing the weak coupling limit. Also, in the trustable area, the phantom phase may appear where the $w$   becomes less than $-1$; but there is no Big-Rip which is indicated by the finite future time divergence  in  scale factors,  energy density, pressures or time derivative of Einstein frame Hubble parameters  \cite{leith2007gauss,nojiri2005properties}.

{The Gauss-Bonnet model coupled to a dynamical scalar field with a non-negative potential on FRW space-times has long been known to have non-singular cosmological solutions by allowing the violations of both the null and the strong energy conditions \cite{kanti1999singularity}.}
Recently, the dark energy scenario has been investigated  
in this model,  where phantom phases have been predicted \cite{nojiri2005gauss}.
The $\Lambda$ term in  Einstein frame effective action \eqref{GBaction} takes the form of a potential of type $V(\phi)=V_0 e^{-\phi}$, where the $V_0$ is assigned to be negative in    $D<26$ dimensional string theory. 
The vanishing $H$-field case  presented  in  \ref{H=0}  and \ref{H=0e} sections, which is described by the similar effective action to that of    Einstein-scalar-Gauss-Bonnet model \cite{nojiri2005gauss} but with a negative potential,  possesses  $w<-1$ phase  in the investigated  $\alpha'$-corrected solutions;  but  the described universe does not seem to reach a  Big-Rip singularity. {Also, it has been shown in \cite{PhysRevD.96.124022,PhysRevD94} that bouncing solutions, which have a connecting phase between a contraction  and an expansion period, are not allowed in the isotropic flat FRW universe in Einstein-scalar-Gauss-Bonnet  model. The presented examples in section \ref{H=0e}  show the similar feature in both isotropic and anisotropic cases, as they have no contraction phase and hence no possibility of appearing a bounce phase.}

The evolution with a non-vanishing homogeneous $H$-field with $H_{012}\neq 0$ has been studied in the
low curvature phase  in  \cite{copeland1994low,copeland1995string}  and in the high curvature phase including the $\alpha'$-corrections in \cite{BF2}, where the potential of the dilaton field or equivalently the $\Lambda$ term has been neglected, assuming the domination of the kinetic terms. However, this assumption made the valid cosmological era of the results to be limited. We have seen that, particularly in the early $\tau$, the $\Lambda$ is significant and cannot be ignored because none of the $R$, $H^2$ or $\dot{\phi}^2$ overcome the $\Lambda$.  
However, in late $\tau$ the curvature and kinetic term of dilaton field may dominate 
where the dynamical effect of the $H$-field becomes negligible. 

{Furthermore, the presented example with non-vanishing $H$-field in section \ref{Hnon0} has no bounce phase in $R^2$ scheme, but  in  Gauss-Bonnet scheme  $\bar{H}$ crosses zero with $\dot{\bar{H}}>0$. Hence, appearing of the bouncing solutions in string inspired  Einstein-scalar-Gauss-Bonnet with $B$-field contribution in the leading and first correction orders seems to be allowed but sensitive to the  chosen RS.}

\appendix
\section{}\label{app1}
In this appendix the explicit forms of the first $\alpha'$-corrections of metric and dilaton field, introduced in \eqref{18}-\eqref{17}, are presented. After performing the integrals in \eqref{46}, \eqref{fi1s}, and \eqref{ini1} with using the \eqref{45} and \eqref{fioa} solutions, we obtain
\begin{widetext}
	\begin{eqnarray}\label{xiI}
	\begin{aligned}
	\xi_i&=-\frac {\Lambda\,{{q_i}}^{2}}{2{n}^{3}} ( \frac{1}{2\, n}\, ( {n
	}^{2}+{\sum q_i^2} )  ( { \cosh^2 (n\tau)}+n^2{\tau}^{2} ) +{\it q_i}\, ( 
	\cosh ( n\tau ) \sinh ( n\tau ) +n\tau ) -
	{\tau}^{2}{n}^{3} ) 
	+{q_i}
	\int\xi_4 \,d\tau+l_i\tau+r_i,
	\end{aligned}
	\end{eqnarray}
	\begin{eqnarray}\label{fi1w}
	\begin{aligned}
	\phi_1=& Q_1\,\tanh(n\tau)+Q_2\,(n\tau \tanh(n\tau)-1)-(n\tanh(n\tau)+\sum q_i)\int \xi_4 \,d\tau+\frac {\Lambda}{4{n}^{4}} \big[- {{\rm e}^{-n\tau}}\sum (q_i^2\, ( {n}^{2}-{\sum q_j^2}
	) \tau\\
	&-4\,{ q_i^3}) \tanh ( n\tau ) n{
		\sum q_i^3}\, ( 2\,  \cosh^2 ( n\tau )  -3 ) +n (4 \sum (n{ q_i^3}+{ q_i^4})+6\,{\sum_{i<j} q_iq_j} ) \tanh ( n\tau ) \tau\\
	&+ (\sum ({n}^{2}{
		q_i^2}+{ q_i^4})+{\sum_{i<j} q_iq_j} )  (  ( 
	\cosh ( n\tau )  ) ^{2}-2 ) -n{\sum q_i^2}\,
	( {n}^{2}-{\sum q_i^2} )  ( n\tau\,\tanh ( n\tau
	) -2 ) \tau\\
	&+\frac {n}{\cosh ( n\tau ) }
	( ( {\sum q_i^2}\, ( {n}^{2}-{\sum q_j^2} ) \tau+4\,{n}^{2}{\sum r_i}-2\,{\sum q_i^3} ) ((n\tau\,\tanh
	( n\tau )-1 ) 
	{{\rm e}^{-n\tau}}-{\frac {n\tau }{\cosh ( n\tau ) }} )) \big]+\varphi_0
	,
	\end{aligned}
	\end{eqnarray}
	where the $l_i$ and $\varphi_0$ are integrating constants and then from \eqref{ini1} we have
		\begin{eqnarray}\label{xi4b0}
	\begin{aligned}
	\xi_4=&\frac {3\Lambda}{4{n}^{3}} \big[ \frac{1}{3}\,n  ( \sum({n}^{2}{ q_i^2}+
	{q_i^4})+{\sum_{i<j} q_i^2q_j^2}  ) \cosh  ( 2\,n\tau  ) +\frac{2
	}{3}\,{n}^{2}{\sum q_i^3}\,\sinh  ( 2\,n\tau  ) -\frac{8}{3}\,n  ( \sum({n
	}^{2}{ q_i^2}+2\,{ q_i^4})\\
&+{\sum_{i\neq j} q_iq_j^3}+\frac{1}{2}{\sum_{i< j} q_i^2q_j^2}
	)  \cosh^2 ( n\tau )-{\frac {20\, }{15}}(
	\sinh  ( n\tau  )   (   ( 5\,{\sum q_i^3}+{\sum_{i\neq j} q_iq_j^2
	}  ) {n}^{2}+ ( \sum q_i )^2\sum q_j \,
	) \cosh  ( n\tau  ))\\
	&-\frac{2}{3}\,n  (\sum ({ q_i^2}\,
	\tau\,{n}^{3}-  ( 2\,{ q_i^3}\,\tau+4\,{  q_i}\,{\sum r_j}
	+3\,{ q_i^2}  ) {n}^{2})        -2\,\tau\,{\sum_{i\neq j} q_iq_j^2}n^2+(\sum q_i^2)^2 (-n\tau+ (2 \tau\,  ( {\sum q_j}
	) -1  ) ) \\
	&+\frac{4}{3}\tanh  ( n\tau  ) {n}^{2
	}  (  \sum( ( { q_i^2}\,\tau+2\,{ r_i}  ) {n}^{2}-{ q_i^3}-\tau\,{ q_i^4})-2\,\tau\,{\sum_{i\neq j} q_i^2q_j^2}  ) +\frac {1 }{ \cosh^2 ( n\tau )}(\frac{2}{3}{n}^{2}  ( \sum(  ( { q_i^2}\,\tau+2\,{ r_i}  ) {n}^
	{2}\\
	&-\tau\,{ q_i^4}+{ q_i^3} ) -2\,\tau\,{\sum_{i\neq j} q_i^2q_j^2}) 
	\cosh  ( n\tau  ) {{\rm e}^{-n\tau}}+\tau\,{n}^{3}  ( 
	(\sum ({ q_i^2}\,\tau+4\,{ r_i}  ) {n}^{2}-\tau\,{ q_i^4}-2\,{ q_i^3})-2\,\tau\,{\sum_{i\neq j} q_i^2q_j^2}  )) \big].
	\end{aligned}
	\end{eqnarray}
\end{widetext}

\section{}\label{App}

With $m=n$ the solution of $\phi_1$ \eqref{fic} reads
\begin{widetext}
\begin{eqnarray}\label{fic1}
\begin{aligned}
\phi_1=
&\frac{1}{n}[\tanh(n\tau)\int (n\tau \tanh(n\tau)-1)\,g_{\phi}(\tau)\,d\tau-(n\tau \tanh(n\tau)-1)\int \tanh(n\tau)\,g_{\phi}(\tau)\,d\tau]-p\int \xi_4 \,d\tau+\varphi_0,
\end{aligned}
\end{eqnarray}
in which   $g_{\phi}$ is simply given by
\begin{eqnarray}\label{gfi}
\begin{aligned}
g_{\phi}(\tau)=&-\frac{2n^2}{\cosh^2 (n\tau)}(\iint (\hat{K}_{3}+\hat{V}_{3})\,d\tau\,d\tau-q_0^{-1}\int \hat{V}_{B}\,d\tau)+\rho.
\end{aligned}
\end{eqnarray}
In the following the explicit forms of the first $\alpha'$-corrections of metric, dilaton field and $H$-field in \eqref{18}-\eqref{17} and \eqref{etaa} are presented for the case of contribution of $H$-field in $\rightarrow$ class, considered in section \ref{Iclass}.
 After performing the integrals of \eqref{ai1}-\eqref{fic} and for the RS of $R^2$, $f=1$, and Gauss-Bonnet, $f=-1$, with $m=n$  we obtain
\begin{eqnarray}\label{aicals1}
\begin{aligned}
\xi_1^{(GB)}=&c_1\,\tanh(n\tau)+c_2\,(n\tau \tanh(n\tau)-1)+(\frac{n}{2}\tanh(n\tau)- q)\int \xi_4 \,d\tau+r_1\\
&+{\frac {\Lambda\,{p}^{2}}{8\,{n}^{2}}}\big[    \tau\,   (    ( {
	n}^{2}\tau-2\,nq\tau-2\,n-4\,q   ) \tanh   ( n\tau   ) -2\,
n+4\,q   )+{
	Li_2}   ( -{{\rm e}^{2\,n\tau}}   ) \tanh   ( n\tau
) \\
&+{\frac {q}{n   }}  ( \cosh   ( 2\,n\tau   )    ) ^{-2} ( n\tau\,   ( {{\rm e}^{-2\,n\tau}}+
1   ) \tanh   ( n\tau   ) -2\,n\tau+{{\rm e}^{-2\,n\tau}}+1
)  +2\,\ln    ( {{\rm e}^{2\,n\tau}}+1   )    
\big]+F(\tau)\mid_{f=-1},
\end{aligned}
\end{eqnarray}
\begin{eqnarray}
\begin{aligned}
\xi_1^{(R^2)}=&c_1\,\tanh(n\tau)+c_2\,(n\tau \tanh(n\tau)-1)+(\frac{n}{2}\tanh(n\tau)- q)\int \xi_4 \,d\tau+r_1\\
&+{\frac {\Lambda}{8\,{n}^{2}}}\big[
( \cosh   ( 2\,n\tau   )    ) ^{-2}\big(-\ln    ( {{\rm e}^{2\,n\tau}}+1   )    (  2\,{n}^{2}-{p}^{2} +  ( n   ( {
	p}^{2}\tau+n   ) \tanh   ( n\tau   ) -{p}^{2}   ) \cosh
( 2\,n\tau   ) \\
&-   ( 2\,  \cosh^2 ( n\tau )-1   ) \tau\,n{p}^{2}\tanh   ( n\tau
)   ) +\frac {  1}{n}( (\frac{1}{2}\,n   ( -2
\,\ln    ( 2   ) {n}^{2}+{n}^{2}   ( {p}^{2}{\tau}^{2}+1
) -2\,{p}^{2}qn{\tau}^{2}+2\,{p}^{2}q\tau   ) \tanh   ( 
n\tau   )\\
& -\frac{1}{2}{p}^{2}   ( {n}^{2}\tau-2\,nq\tau-2\,q   ) 
) {{\rm e}^{-2\,n\tau}}+\tau\,n   ( n   ( \frac{1}{2}\,n{p}^{2}
\tau-{p}^{2}q\tau+{n}^{2}   ) \tanh   ( n\tau   ) -\frac{1}{2}\,{p}
^{2}   ( n-2\,q   )    ) {{\rm e}^{2\,n\tau}}\\
&+n   ( -
\ln    ( 2   ) {n}^{2}+{n}^{3}\tau+   ( {p}^{2}{\tau}^{2}+\frac{1}{2} ) {n}^{2}-2\,{p}^{2}qn{\tau}^{2}+{p}^{2}q\tau   ) \tanh
( n\tau   ) +2\,{n}^{4}\tau-{n}^{2}{p}^{2}\tau+2\,{n}^{3}
\ln    ( 2   ) -{n}^{3}+{p}^{2}q)
\big)\\
&-\tanh   ( n\tau   ) {p}^{2}   ( -{Li_2}   ( -{
	{\rm e}^{2\,n\tau}}   ) +\tau\,   ( n\tau\,   ( n-2\,q
) +2\,n+4\,q   )    )
\big]+F(\tau)\mid_{f=1},
\end{aligned}
\end{eqnarray}
where the $F(\tau)$ is given by
\begin{eqnarray}
\begin{aligned}
F(\tau)=&\frac{\Lambda}{32\,n^3}\big(
\tanh(n\tau)\big(8\,{n}^{3}\ln  ( {{\rm e}^{2\,n\tau}}+1 ) \tau\, ( 
( 3\,{n}^{2}+8\,{p}^{2}+4\,{q}^{2} )f +{n}^{2}+2\,{p}^{2}+4
\,{q}^{2} ) -4\,n\tau\, (  ( n\tau+2
) {n}^{2} ( 3\,{n}^{2}+8\,{p}^{2}\\
&+4\,{q}^{2} ) f+
\tau\,{n}^{5}+3\,{n}^{4}+2\,\tau\, ( {p}^{2}+2\,{q}^{2} ) {
	n}^{3}+ ( 4\,{p}^{2}\tau\,q+6\,{p}^{2} ) {n}^{2}+8\,{p}^{2}
{q}^{2}+16\,{q}^{4} ) +4\,{n}^{2}{ Li_2} ( -{{\rm e}^{2\,n\tau}}
)  ( f ( 3\,{n}^{2}\\
&+8\,{p}^{2}+4\,{q}^{2} ) +{n
}^{2}+2\,{p}^{2}+4\,{q}^{2} ) -\sinh ( 2\,n\tau )  ( 48
\,{q}^{4}+ ( 16\,\tau\,{n}^{3}+24\,{p}^{2} ) {q}^{2}-16\,{n
}^{2}{p}^{2}\tau\,q+3\,{n}^{4}+6\,{n}^{2}{p}^{2} )\\
& +2\,n\cosh
( 2\,n\tau )  (  ( {n}^{4}+2\,{n}^{2}{p}^{2}+8\,
{p}^{2}{q}^{2}+16\,{q}^{4} ) \tau-12\,n{q}^{2}+12\,{p}^{2}q
) +{\frac {{n}^{4} ( 10\,f+13 )  ( 2\,n\tau-{
			{\rm e}^{-2\,n\tau}}-1 ) }{  \cosh ^{2}( n\tau ) 
		}}
\big)\\
&-(n\tau\tanh(n\tau)-1)\big(
2\,  \cosh ^{2} ( n\tau )   (  ( {n}
^{2}-4\,{q}^{2} ) ^{2}+2\,{n}^{2}{p}^{2}+8\,{p}^{2}{q}^{2}
) +16\,\sinh ( n\tau ) {p}^{2}q\cosh ( n\tau
) n\\
&-16\,{p}^{2}\tau\,q{n}^{2}+4\,{n}^{2}\ln  ( \cosh
( n\tau )  )  (  ( 3\,{n}^{2}+8\,{p}^{2}+
4\,{q}^{2} )f +{n}^{2}+2\,{p}^{2}+4\,{q}^{2} ) -{n}^{4}
 \tanh^{2}  ( n\tau )  ( 10\,f+13
) 
\big)\big).
\end{aligned}
\end{eqnarray}
Also, we have
\begin{eqnarray}
\begin{aligned}
\xi_2=& \xi_1+2\, q\int \xi_4 \,d\tau+{\frac {\Lambda\,{p}^{2}q  }{2\,{n}^{3}}} ( \sinh  ( n\tau  ) \cosh
( n\tau  ) +n\tau  )+r_2
,
\end{aligned}
\end{eqnarray}
\begin{eqnarray}
\begin{aligned}
\xi_3=p\int \xi_4 \,d\tau+{\frac {{p}^{2}\Lambda}{4\,{n}^{4}}}\big[&  ( \frac{3}{2}{n}^{2}+{p}^{2}+2\,{q}^{2}  )   \cosh^2 ( n\tau )+2\,\sinh  ( n\tau  ) \cosh  ( 
n\tau  ) np\\
&+  ( \tau\,  ( { f}+\frac{1}{2}  ) {n}^{2}+{p
}^{2}\tau+2\,{q}^{2}\tau+2\,p  ) {n}^{2}\tau
\big]+r_3,
\end{aligned}
\end{eqnarray}

\begin{eqnarray}
\begin{aligned}
\phi_1^{(GB)}=&U(\tau)\mid_{f=-1} -p\,\int \xi_4 \,d\tau+\frac{\Lambda p^3}{8\, n}\big(  \tanh(n\tau)(-n{\tau}^{2}-2\,\tau+2\,\tau\ln  ( 1+{{\rm e}^{2\,n\tau}} ) +{\frac {1}{n}}{
	Li_2} ( -{{\rm e}^{2\,n\tau}} ) 
) \\
&-\frac{2}{n}(n\tau\tanh(n\tau)-1)\ln(\cosh(n\tau))
\big)+\varphi_0,
\end{aligned}
\end{eqnarray}
\begin{eqnarray}
\begin{aligned}
\phi_1^{(R2)}&=U(\tau)\mid_{f=1} -p\,\int \xi_4 \,d\tau+\frac{\Lambda }{8\, n^2}\big(  \tanh(n\tau)({ ( \cosh ( n\tau )  ) ^{-2}} n ( 2\,{n}^{2}{\tau}^{2}+4\,n\ln  ( 2 ) \tau+1
/2+n\tau\, ( 2\,\sinh ( 2\,n\tau )-1 \\
&+2\,\ln  ( 2
) {{\rm e}^{2\,n\tau}} ) +\frac{1}{2}{{\rm e}^{-2\,n\tau}}
) +\ln 
( 1+{{\rm e}^{2\,n\tau}} )  ( 2\,{p}^{2}\tau-{\frac {3n  }{4\cosh^{2}  ( n
		\tau )  }}( 
2\,n\tau+\sinh ( n\tau )  ) ) +{\frac {{p}^{2}}{n}}{
	Li_2} ( -{{\rm e}^{2\,n\tau}} ) -n\ln  ( 
2 ) \\
&-\frac {{n}^{
		2}\tau\,{{\rm e}^{n\tau}}}{2\cosh ( n\tau ) }+ ( -n{\tau}^{2}-2\,\tau ) {p}^{2}+{n}^{2}\tau
) +(n\tau\tanh(n\tau)-1)({\frac { 2({n}^{2}  - \cosh^{2} ( n\tau )  {p}^{
			2}) \ln  ( \cosh ( n\tau )  ) +{n}^{2}}{n \cosh  ^{2}( n\tau )  }}
)
\big)+\varphi_0,
\end{aligned}
\end{eqnarray}
where the $U(\tau)$ is given by
\begin{eqnarray}
\begin{aligned}
U(\tau)=&\frac{\Lambda}{32\,n^3}\big(
\tanh(n\tau)\big(
\tau\, ( 11\,{n}^{5}\tau+ ( -8\,\tau+18 ) {n}^{4}+2\,
\tau\, ( {p}^{2}+58\,{q}^{2}+2 ) {n}^{3}+ (  ( 24
\,{p}^{3}-16\,{p}^{2}-32\,{q}^{2}+8\,p ) \tau+44\,{p}^{2}\\
&+104\,{
	q}^{2}+4 ) {n}^{2}+ ( -4\,{p}^{4}\tau-8\,{p}^{2}{q}^{2}\tau
+16\,{p}^{3} ) n+24\,{p}^{4}+80\,{p}^{2}{q}^{2}+64\,{q}^{4}+16\,
{q}^{2} ) -n{ Li_2} ( -{{\rm e}^{2\,n\tau}} ) 
(  ( 18\,{n}^{2}+36\,{p}^{2}\\
&+24\,{q}^{2} ) f+11\,{n}^
{2}+4\,{p}^{2}+116\,{q}^{2}+4 ) +\frac {1}{4  \cosh ^{2}
		( n\tau )  } ( 4\,f{n}^{3} ( 4\,n{p}
	^{2}{\tau}^{3}+2\,{p}^{2}{\tau}^{2}-34\,n\tau+17 ) +8\,n{p}^{2}
	( {n}^{2}+2\,{p}^{2}\\
	&+4\,{q}^{2} ) {\tau}^{3}+4\,{p}^{2}
	( {n}^{2}+8\,np+2\,{p}^{2}+4\,{q}^{2} ) {\tau}^{2}-4\,
	( { {35\,{n}^{3}}}+2\,{p}^{3} ) \tau+70\,{n}^{2}+
	( 8\,f ( {p}^{2}{\tau}^{2}+17/2 ) {n}^{3}+4\,{p}^{2}
	(  ( {n}^{2}+2\,{p}^{2}\\
	&+4\,{q}^{2} ) \tau-8\,p
	) \tau\,n+70\,{n}^{3} ) {{\rm e}^{-2\,n\tau}} ) -2
\,{n}^{2}\tau\,\ln  ( 1+{{\rm e}^{2\,n\tau}} )  ( f(18\,
{n}^{2}+36\,{p}^{2}+24\,{q}^{2})+11\,{n}^{2}+4\,{p}^{2}+116\,{q}^{2}
+4 ) \\
&+2\, (  ( -4\,{p}^{4}-8\, ( {n}^{2}+2\,{q}^{
	2} ) {p}^{2}+{n}^{4}+ ( 32\,{q}^{2}+1 ) {n}^{2}-16\,{
	q}^{4}-4\,{q}^{2} ) \tau+6(4\,{p}^{3}-{n}^{2}-2\,{p}^{2}-4\,
{q}^{2}+p) ) \cosh ( 2\,n\tau ) \\
&+\frac {1 }{n} ( 
( 8\,\tau-3 ) {n}^{4}+ ( -32\,{p}^{3}\tau+ 8( 2
\,\tau+3 ) {p}^{2}-8\,p\tau-3+32 ( \tau-3 ) {q}^{
	2} ) {n}^{2}+12\,({p}^{4}+{q}^{2} ( 4\,{p}^{2}+4\,{q}^{2}+
1 ) ) ) \sinh ( 2\,n\tau )
\big)\\
&-(n\tau\tanh(n\tau)-1)\big(-2\,n\ln  ( 1+{{\rm e}^{2\,n\tau}} )  (  ( 18\,{n
}^{2}+36\,{p}^{2}+24\,{q}^{2} ) f+11\,{n}^{2}+4\,{p}^{2}+116\,{q
}^{2}+4 ) +\tau\, ( f{n}^{2} ( 9\,{n}^{2}\tau\\
&+64\,{p}^
{2}+48\,{q}^{2} ) +22\,{n}^{4}-16\,{n}^{3}+ ( 4\,{p}^{2}+
232\,{q}^{2} ) {n}^{2}+ ( 48\,{p}^{3}-32\,{p}^{2}-64\,{q}^{
	2}+16\,p ) n-8\,{p}^{4}-16\,{p}^{2}{q}^{2} ) \\
&+{ ( \cosh ( n\tau )  ) ^{-2}} 
		4( n{\tau}^{2}+\tau ) {p}^{4}+4 ( 2\,n\tau-1
		) {p}^{3}+2\, ( n\tau+1 )  (  ( 2\,f+1
		) {n}^{2}+4\,{q}^{2} ) \tau\,{p}^{2}+ ( -34\,f-35
		) {n}^{3}+4\, ( {p}^{2}\tau\\
		&-p+ (  ( f+1/2
		) {n}^{2}+2\,{q}^{2} ) \tau ) {p}^{2}{{\rm e}^{-2\,
				n\tau}}+{\frac 
	{2 }{n}}(\cosh ( 2\,n\tau )  ( {n}^{4}+ ( -8\,{p}^{2}+32
\,{q}^{2}+1 ) {n}^{2}-16\,{q}^{4}-4((4 {p}^{2}+
) {q}^{2}+{p}^{4}) ))\\
&+8\, ( -4\,{p}^{3}+{n}^{
	2}+2\,{p}^{2}+8\,{q}^{2}-p ) \sinh ( 2\,n\tau )
\big)
\big),
\end{aligned}
\end{eqnarray}
\begin{eqnarray}
\begin{aligned}
\eta^{(GB)}={\frac {b{p}^{2}\Lambda\, }{8\,{n}^{3}}}  ( \sinh  ( n\tau  ) 
\cosh  ( n\tau  ) +{n\tau}  ),
\end{aligned}
\end{eqnarray}
\begin{eqnarray}\label{etar2}
\begin{aligned}
\eta^{(R^2)}={\frac {b\Lambda}{4\,n}  ( -{n}^{2}{\tau}^{2}-2\,\ln   ( 2
	) n\tau-{ Li_2}  (-{{\rm e}^{2\,n\tau}}  ) 
	+{\frac {{p}^{2}\sinh  ( 2\,n\tau  ) }{4\,{n}^{2}}}+\frac{1}{2}{
		\frac {{p}^{2}\tau}{n}}  ) }.
\end{aligned}
\end{eqnarray}
Also, the equation \eqref{91} gives
\begin{eqnarray}
\begin{aligned}
\xi_4^{(GB)}=&\frac {-\Lambda}{16\,{n}^{2}(2\,{n}^{2}-2\,{p}^{2}-4\,{q}^{2})} \big[\ln   ( {{\rm e}^{2\,n\tau}}+1  )   ( 7\,{n}^{2
}+40\,{p}^{2}-92\,{q}^{2}-4  ) {n}^{2}-\tau\,{n}^{2}  ( 7\,{n}^{4}+8\,{n}^{3}+  ( 38\,{p}^{2}-92\,{q
}^{2}-4  ) {n}^{2}
\\
&+  ( -24\,{p}^{3}+8\,  ( -q+2  ) {
	p}^{2}-8\,p+32\,{q}^{2}  ) n+4\,{p}^{4}+8\,{p}^{2}{q}^{2}
) +  ( {n}^{4}+  ( -8
\,{p}^{2}+32\,{q}^{2}+1  ) {n}^{2}-4\,{p}^{4}-
16\,{q}^{4}-4\,{q}^{2}\\
&-16\,{p}^{2}{q}^{2}  ) \cosh  ( 2\,n\tau  )+  ( 
4\,{n}^{3}+  ( -16\,{p}^{3}+8\,{p}^{2}+16\,{q}^{2}-4\,p  ) n
) \sinh  ( 2\,n\tau  ) +{\frac {1 }{2\cosh^2 ( n\tau )}}(-2\,n{p}^{2}
( {n}^{2}\tau\\
&-2\,{p}^{2}\tau-4\,{q}^{2}\tau+2\,p-8\,q  ) {
	{\rm e}^{-2\,n\tau}}+  ( 12\,{n}^{4}+  ( 16\,{p}^{2}+224\,{q}^
{2}+8  ) {n}^{2}-64\,{q}^{4}-32\,{q}^{2}  )   \cosh^4 ( n\tau )+32\,  ( -2\,{p}^{3}\\
&+{n}^{2}+
( q+2  ) {p}^{2}-p+4\,{q}^{2}  ) n\sinh  ( n\tau
)   \cosh^3 ( n\tau )+  ( 6\,{
	n}^{4}+  ( -48\,{p}^{2}-216\,{q}^{2}-8  ) {n}^{2}-48\,{p}^{4}
\\
&-96\,{p}^{2}{q}^{2}  )   \cosh^2 ( n\tau )-16\,n{p}^{2}\sinh  ( n\tau  )   ( {n}^{2}
\tau-2\,{p}^{2}\tau-4\,{q}^{2}\tau+2\,p+q  ) \cosh  ( n\tau
)+n  (   (6 {p}^{2}{\tau}^{2}-7  ) {n}^{3}\\
&-2\,{n}^{2}\tau\,{p}^{2}-12\,{p}^{2}\tau\,  ( {p}^{2}\tau+2\,{q}^{2}
\tau+2\,p+4\,q  ) n+4\,{p}^{2}  ( {p}^{2}\tau+2\,{q}^{2}
\tau-p+4\,q  )   ))+2(  \cosh^{2} ( n\tau )   (  ( {n}^{2
}-4\,{q}^{2} ) ^{2}\\
&+2\,{p}^{2} ( {n}^{2}+4\,{q}^{2}
)  ) +8\,\cosh ( n\tau ) n{p}^{2}q\sinh
( n\tau ) -8\,{p}^{2}\tau\,q{n}^{2}+2\,{n}^{2}\ln  ( 
\cosh ( n\tau )  )  ( -2\,{n}^{2}-6\,{p}^{2}
) \\
&-{3}/2{n}^{4}  \tanh^{2} ( n\tau )  
-2\,{ {{n}^{4}}{ ( \cosh ( n\tau )  ) ^{-2}}
}
)
\big],
\end{aligned}
\end{eqnarray}
\begin{eqnarray}\label{aicals4}
\begin{aligned}
\xi_4^{(R^2)}&=\frac {-\Lambda(\cosh  ( n\tau   )    ) ^{-2}}{1184(\,{n}^{2}-2\,{p}^{2}-4\,{q}^{2}) } \big[-592\,\ln   ( 
\cosh  ( n\tau  )   ) {n}^{2}-  ( \cosh  ( 2\,n\tau  )   ( 29\,{n}^{2}-112\,{p}^{2
}+140\,{q}^{2}+4  ) +333\,{n}^{2}\\
&-112\,{p}^{2}+140\,{q}^{2}+4
) \ln   ( {{\rm e}^{2\,n\tau}}+1  )+\frac { 1}{2{n}^{2}}( ( 58\,{
	n}^{5}\tau+  ( -16\,\tau+2  ) {n}^{4}+  ( 84\,{p}^{2}\tau
+280\,{q}^{2}\tau+8\,\tau  ) {n}^{3}\\
&+  ( 48\,{p}^{3}\tau+
-16\,( 2\tau+1  ) {p}^{2}+16\,p\tau+2+ 64\, ( -\tau+1
) {q}^{2}  ) {n}^{2}-8\,{p}^{4}-32\,{p}^{2}{q}^{2}-32\,{q}
^{4}-8\,{q}^{2}  ) \cosh  ( 2\,n\tau  ) \\
&+  ( {n}^{4}
-4\,{n}^{3}+  ( -8\,{p}^{2}+32\,{q}^{2}+1  ) {n}^{2}-4\,{p}^{
	4}-16\,{p}^{2}{q}^{2}-16\,{q}^{4}-4\,{q}^{2}  ) \cosh  ( 4\,n
\tau  ) +8\,  ( -{p}^{4}\tau-4\,{p}^{3}+{n}^{2}\\
&+  2\,( -{
	q}^{2}\tau+1  ) {p}^{2}-p+4\,{q}^{2}  ) n\sinh  ( 2\,n
\tau  )+4\,n  ( -4\,{p}^{3}+{n}^{2}+2\,{p}^{2}+4\,{q}^{2}-p
) \sinh  ( 4\,n\tau  ) -8\,  ( 17\,{n}^{2}\tau-37
\,nq\tau\\
&+p-37\,q  ) n{p}^{2}{{\rm e}^{-2\,n\tau}}-148\,\tau\,{n}^
{2}{p}^{2}  ( n-2\,q  ) {{\rm e}^{2\,n\tau}}+  ( 888\,{n}
^{4}+  ( 1184\,{p}^{2}+16576\,{q}^{2}+592  ) {n}^{2}-4736\,{q
}^{4}\\
&-2368\,{q}^{2}  )   \cosh^4 ( n\tau )+2368\,  ( -2\,{p}^{3}+{n}^{2}+  ( q+2  ) {p
}^{2}-p+4\,{q}^{2}  ) n\sinh  ( n\tau  )   \cosh^{3}
( n\tau  )   +  ( -3108\,{n}^{4}\\
&+  ( -
3552\,{p}^{2}-20720\,{q}^{2}-592  ) {n}^{2}-3552\,{p}^{4}-7104\,{
	p}^{2}{q}^{2}  )   \cosh^2 ( n\tau )
 +1184\,n\sinh  ( n\tau  )   ( 3\,{n}^{2}\tau+2\,{p}^{2}
\tau+4\,{q}^{2}\tau\\
&-2\,p-q  ) {p}^{2}\cosh  ( n\tau  ) +
608\,{n}^{4}\ln   ( 2  ) +666\,{n}^{5}\tau+  ( -1764\,{p}
^{2}{\tau}^{2}-16\,\tau+7115  ) {n}^{4}+  ( -224\,{p}^{2}\tau
+280\,{q}^{2}\tau+8\,\tau  ) {n}^{3}\\
&+  ( -1176\,{p}^{4}{\tau}
^{2}-2304\,{p}^{3}\tau+  ( -2352\,{q}^{2}{\tau}^{2}-1184\,q\tau-32
\,\tau-8  ) {p}^{2}+16\,p\tau+1+  ( -64\,\tau+32  ) {q}^
{2}  ) {n}^{2}\\
&-8\,{p}^{2}  ( -37\,q+p  ) n-4(\,{p}^{4}+4
\,{p}^{2}{q}^{2}+4{q}^{4}+{q}^{2}))+148\,n ( {\frac { 1 }{{2 n}^{3}}} \cosh^{4} ( n\tau ) 
 (  ( {n}^{2}-4\,{q}^{2} ) ^{2}+2{p}^{2}(\,{n}^{2
}+4{q}^{2}) )-n\\
&+{\frac { 1}{{n}^{2}} \cosh^{2}
	( n\tau )   ( 4\,\sinh ( n
	\tau ) {p}^{2}q\cosh ( n\tau ) -4\,{p}^{2}\tau\,qn+2n
	\ln  ( \cosh ( n\tau )  )  ( 2{n}^{2}+5
	\,{p}^{2}+4\,{q}^{2} ) -{\frac {23\,{n}^{2} }{4}} \tanh^{2}
	( n\tau )   ) }) 
\big].
\end{aligned}
\end{eqnarray}
\end{widetext}


\bibliography{mybib1}

\end{document}